\documentclass[aps,prd,a4paper,preprintnumbers,amsmath,amssymb,superscriptaddress,twocolumn,floatfix,showpacs]{revtex4}
\usepackage{amssymb}
\usepackage{float}
\usepackage[dvips]{graphics}
\usepackage[pdftex]{graphicx}

\usepackage{amssymb}
\newcommand{\dV}{\sqrt{-g}{\rm d}^4 x\,}
\newcommand{\kap}{\kappa_{4}}

\newcommand{\etal}{\emph{et al.}}

\newcommand{\be}{\begin{eqnarray}}
\newcommand{\ee}{\end{eqnarray}}
\newcommand{\ba}{\left( \begin{array}{ccc}}
\newcommand{\ea} {\end{array} \right)}
\newcommand{\bv}{\left( \begin{array}{c}}
\newcommand{\ev} {\end{array} \right)}

\newcommand{\dd}{\mathrm{d}}

\newcommand{\gsimm}{\raise.3ex\hbox{$>$\kern-.75em\lower1ex\hbox{$\sim$}}}
\newcommand{\lsimm}{\raise.3ex\hbox{$<$\kern-.75em\lower1ex\hbox{$\sim$}}}

\begin{document}
\bibliographystyle{unsrt}


\title{The Dilaton and Modified Gravity}

\author{Philippe Brax}
\affiliation{Institut de Physique Th\'eorique, CEA, IPhT, CNRS, URA 2306,
  F-91191Gif/Yvette Cedex, France}
\email{philippe.brax@cea.fr}

\author{Carsten van de Bruck}
\affiliation{Department of Applied Mathematics,
  University of Sheffield Hounsfield Road, Sheffield S3 7RH, United
  Kingdom}
\email{c.vandebruck@sheffield.ac.uk}

\author{Anne-Christine Davis}
\affiliation{Department of Applied Mathematics and
  Theoretical Physics, Centre for Mathematical Sciences, Cambridge CB3
  0WA, United Kingdom}
\email{a.c.davis@damtp.cam.ac.uk}

\author{Douglas Shaw}
\affiliation{Queen Mary University of London, Astronomy Unit,
Mile End Road, London E1 4NS, United Kingdom}
\email{d.shaw@qmul.ac.uk}

\date{\today}

\begin{abstract} We consider the dilaton in the strong string coupling limit  and elaborate on the original idea of Damour and Polyakov whereby the dilaton coupling to matter has a minimum with a vanishing value at finite field-value. Combining this type of coupling with an exponential potential, the effective potential of the dilaton becomes matter density dependent. We study the background
cosmology, showing that the dilaton can play the role of dark energy. We also analyse
the constraints imposed by the absence of violation of the equivalence principle. Imposing these constraints and assuming that the dilaton plays the role of dark energy, we consider the
consequences of the dilaton on large scale structures and in particular the behaviour of the slip functions and the growth index at low redshift.
\end{abstract}

\maketitle

\section{Introduction}
The observed late time acceleration of the Universe has no clear theoretical explanation yet. One of the putative candidates is dark energy whereby a scalar field with a flat enough potential provides the potential energy leading to the accelerating phase (for reviews and references see e.g. \cite{various}). Although this seems like a natural scenario, it is fraught with difficulties. The most important one is certainly the absence of a complete understanding of the role of quantum corrections in such models. This is akin to the hierarchy problem of Higgs physics, albeit even more serious due to the stringent phenomenological constraints that acceleration imposes on the dark energy potential. In particular, both the vacuum energy and the mass of the scalar field must be minute. The former is nothing but a reformulation of the cosmological constant problem whereas the latter prescribes that there should exist a new fifth force complementing the gravitational interaction at very large scales. Reconciling such a long range force with local experiments of gravity on earth and in the solar system is a difficult task. Three types of mechanisms can be generically invoked. The first one appears in the DGP \cite{DGP} modification of gravity where the Vainshtein \cite{Vainshtein} effect is present locally. In these models, gravity is modified on large scales and preserved close to massive bodies due to the non-linearities of the scalar field kinetic terms. The shielding of the scalar field by massive bodies is also a feature of chameleon models \cite{chameleon1,chameleon2,chameleon3} (and therefore of $f(R)$-gravity \cite{f(R)chameleon}) where the mass of the scalar field becomes environmentally dependent. This leads to a thin shell effect preventing any deviations from Newton's law in the vicinity of massive objects.

Another mechanism has been advocated in a string theoretic context: the Damour-Polyakov effect \cite{Damour:1994zq}. Considering the string dilaton in the strong coupling regime, it turns out that no violation of general relativity would be observed provided the coupling of the dilaton to matter were driven to zero by the cosmological expansion. This results stands when no potential is taken into account for the dilaton. It turns out that in the strong coupling regime, one expects that the dilaton will have an exponentially decreasing potential akin to the ones used to describe dark energy. This was noticed in \cite{ven1, ven2, ven3} and \cite{ven1} proposed a scenario whereby quintessence was the runaway dilaton, though the couplings to matter were neglected here. Later this was
modified and  gravitational tests evaded provided the coupling of the dilaton to matter vanishes for an infinitely large dilaton. This is equivalent  to the Damour-Polyakov mechanism with a minimum at infinity.

In this paper, we will focus on the original Damour-Polyakov setting where the coupling vanishes for a finite value of the dilaton while keeping an exponentially runaway dilaton potential. In this case, the potential term tends to displace the dilaton from the minimum with no coupling to matter. The fifth force constraints would not be evaded anymore. In fact this result only stands when the string and Planck scales are of the same order of magnitude. Provided the string scale is lower that the Planck scale by a few orders of magnitude, we find that the Damour-Polyakov mechanism is at play albeit only locally where matter densities are large. This allows one to evade solar system constraints on gravity. This environmentally dependent Damour-Polyakov mechanism implies that there exists a fifth force whose manifestation is prominent on galaxy cluster scales. This prompts the interesting possibility that one could probe and measure dilatonic modifications of gravity with future large scale surveys.

This paper is organised as follows. In Section 2, we will recall the main properties of dilaton models in the strong coupling regime. Then, in Section 3 we will study the local tests and the constraints they impose on dilatonic models and investigate the chameleon mechanism as well as the Damour-Polyakov mechanism. In Section 4, we present the environmentally dependent Damour-Polyakov mechanism and discuss the cosmological evolution as well as the the local constraints. In Section 5 we will focus on the  consequences for large scale structures in the universe. Our conclusions can be found in Section 6.

\section{Dilaton and Modified Gravity}

\subsection{Dilaton models}

Our starting point is the low energy, gravi-dilaton string-frame effective action, including dilaton-dependent corrections, as given in \cite{ven1,ven2} (see also \cite{Green:1987sp}):

\be
\mathcal{S} &=& \int \sqrt{-\tilde{g}}\dd^4 x\,\left[ \frac{e^{-2\psi(\phi)}}{2 l_{\rm s}^2}\tilde{R}  + \frac{Z(\phi)}{2l_{\rm s}^{2}}(\tilde{\nabla}^2 \phi) - \tilde{V}(\phi)\right] \nonumber \\ &&
+ \mathcal{S}_{\rm m}\left(\Psi_{i},\tilde{g}_{\mu \nu}; g_{i}(\phi) \right).\nonumber
\ee

Here $l_{\rm s}$ is the string length-scale, $\Psi_{i}$ are the matter fields; $\tilde{R}$ is the Ricci scalar curvature of $\tilde{g}_{\mu \nu}$. The $g_{i}(\phi)$ represent the `constants' of nature such as the gauge coupling constants, which are now dilaton dependent.

We move to the arguably more physically transparent Einstein frame, by defining $\tilde{g}_{\mu \nu} = A^2(\phi) g_{\mu \nu}$ where $A(\phi) = l_{\rm s} e^{\psi(\phi)}/ \kap$; $\kap^2 = 8\pi G_N$.  We are also free to rescale $A(\phi)$ by a constant factor, and so fix its definition by requiring $A(\phi_{0}) = 1$, where $\phi_0$ is approximately the value of $\phi$ today.  We let $c_{1} \equiv  l_{\rm s}/\kappa_4 = \exp(-\psi(\phi_0)))$.  The Einstein frame action then becomes:
\be
\mathcal{S} &=& \int \dV \left(\frac{R(g)}{2\kap^2} - \frac{k^{2}(\phi)}{\kap^2}(\nabla \phi)^2 - V(\phi)\right) \label{action} \\ &&+ \mathcal{S}_{\rm m}\left(\Psi_{i}, A^2(\phi)g_{\mu \nu}; \phi\right), \nonumber
\ee
where $k^{2}(\phi) = 3\beta^2(\phi)-A^2(\phi) Z(\phi)/2c_1^2$, where $\beta(\phi) = (\ln A)_{,\phi}$, and $V(\phi) = A^4(\phi)\tilde{V}(\phi)$.

We will additionally assume that $e^{-\phi_0} \ll 1$ so that we are in
the strong-coupling limit. In this limit $(\phi \rightarrow \infty)$,
we assume (see \cite{brustein} and references therein):
\be
\tilde{V}(\phi) &\sim& \tilde{V}_{0}e^{-\phi} + \mathcal{O}(e^{-2\phi}), \nonumber \\
Z(\phi) &\sim& -\frac{2c_{1}^2}{\lambda^2} + b_{Z}e^{-\phi} + \mathcal{O}(e^{-2\phi}), \nonumber \\
g_{i}^{-2} &\sim&  \bar{g}_{i}^{-2} + b_{i} e^{-\phi} + \mathcal{O}(e^{-2\phi}). \nonumber
\ee
We will assume that $e^{-\phi_0}$ is sufficiently small that these asymptotic expansions are valid. In \cite{ven1} these potential and couplings were derived and it was showed that the dilaton could in principle act as dark energy. Typically one expects $b_{Z}\sim O(1)$ and $b_{i}\sim O(1)$. Natural values for $\lambda$ range from  $\lambda \sim O(1)$ to $O(c_{1}) = M_{\rm pl}/M_{\rm s}$, and generally $c_{1} \gg 1$.   Using $e^{-\phi_0} \ll 1$, we have
\be
k(\phi) \approx \lambda^{-1} \sqrt{1+ 3\lambda^2 \beta^2(\phi)}.
\ee
We define $\varphi$ by $\dd \varphi = k(\phi) \dd \phi$.  The field equation for $\phi$ (or equivalent $\varphi$) is then given by
\be
\square \varphi &=& \frac{\kap^2}{2 k(\phi)} \left[-V(\phi) - \beta(\phi) \left(A^4(\phi)\tilde{T}_{\rm  m}-4V(\phi)\right) \right. \\ && \left. - \sum_{i} A(\phi)\beta_{i}(\phi)(\phi) S_{i} \right],\nonumber
\ee
where
\be
\tilde{T}^{\mu \nu}_{\rm m} &=& \frac{2}{\sqrt{-\tilde{g}}} \frac{\delta \mathcal{S}_{\rm m}}{\delta \tilde{g}_{\mu \nu}}, \nonumber \\
A(\phi)\beta_{i} &=& \frac{\delta \ln g_{i}}{\delta \phi} \sim -\frac{g_{i}^2(\phi) b_{i}e^{-\phi}}{2}, \nonumber \\
 S_{i} &=& \frac{\delta \mathcal{S}_{\rm m}}{\delta \ln g_{i}}, \nonumber
\ee
and $\tilde{T}_{\rm m} = \tilde{T}^{\mu \nu}_{\rm m}\tilde{g}_{\mu \nu}$. We have defined  the conserved matter density in the Einstein frame $T_m= g_{\mu \rho} T_{\rm m}^{\rho \nu} = A^{3}\tilde T_m $. Typically the $S_{i}$ are $\mathcal{O}(T_{\rm m})$ or smaller.  We assume that, near $\phi_0$, $e^{-\phi}$ is sufficiently small that the composition dependent couplings $\beta_{i} \propto e^{-\phi}$ are suppressed to be much smaller than the universal coupling $\beta$.

The field equation for $\phi$ then simplifies to
\be
\square \varphi \approx \frac{\kap^2}{2 k(\phi)} \left[-V(\phi) - \beta(\phi)\left(A(\phi) T_{\rm  m} - 4 V(\phi)\right) \right].
\ee

We may think of $\varphi$ as feeling an effective potential $V_{\rm eff}(\varphi; T_{\rm m})$ defined by
\be
\square \varphi = \frac{\kap^2}{2} V_{{\rm eff}, \varphi}(\varphi; T_{\rm m}) = \frac{\kap^2}{2k(\phi) } V_{{\rm eff}, \phi}(\phi; T_{\rm m}). \nonumber
\ee
It follows that
\be
 V_{{\rm eff}}(\varphi; T_{\rm m}) =  V_{0}A^4(\phi) e^{-\phi} - A(\phi) T_{\rm m}~.
\ee
With pressureless matter, $T_{\rm m} = -\rho_{\rm m}$, and this effective potential is minimized when $\phi=\phi_{\rm min}(\rho_{\rm m})$ which is given by:
\be
\beta(\phi_{\rm min}) = \frac{V(\phi_{\rm min})}{A(\phi_{\rm min})\rho_{\rm m} + 4V(\phi_{\rm min})}.
\ee
This implies that $\beta(\phi_{\rm min}) \leq 1/4$.   Cosmologically, today if $V(\phi)$ is responsible for the late-time acceleration of the Universe, we have $A\rho_{\rm m}/V \approx 0.27/0.73 \approx 0.37$, and so cosmologically $\beta \approx 0.23$ today.

We have used the fact  that, cosmologically, for pressure-less matter, $T_{\rm m} = \rho_{\rm m}(t) = \rho_{\rm m0} a_0^3/a^3$ does not depend explicitly on $\phi$. We define $m_{\varphi}$ to be the effective mass of small perturbations in $\varphi$
\be
m_{\varphi}^2(\phi;T_{\rm m}) &=& V_{{\rm eff},\varphi\varphi}(\varphi_{\rm min};T_{\rm m}) \\ &=& k^{-2}(\phi)\left[V_{{\rm eff},\phi\phi}(\varphi_{\rm min};T_{\rm m})\right. \\ && \left. - \frac{1}{2}(\ln k^2)_{,\phi}V_{{\rm eff},\phi}\right]. \nonumber
\ee
Thus, at the minimum of the effective potential
\be
m_{\varphi}^2(\phi_{\rm min};T_{\rm m}) &=& \frac{\kap^2 \beta_{,\phi \phi}}{2k^2(\phi)}(A \rho_{\rm m}+4V) \\ &&+ \frac{\kap^2 \beta (1-3\beta) A \rho_{\rm m}}{2k^2(\phi)}. \nonumber
\ee
Cosmologically, $\kap^2 A\rho_{\rm m} = 3\Omega_{\rm m}H^2$ and $\kap^2 V = 3\Omega_{\Lambda}H^2$, and we have $\beta \leq 1/4$ at the minimum of $V_{\rm eff}$. Generally then, unless $\beta_{,\phi\phi} \gg 1$, we have $m_{\varphi} \sim O(H)$, leading to a new long-range force over cosmological scales.

\subsection{Dilatonic modification of gravity}

Over length scales less than $\lambda_{\varphi} = m_{\varphi}^{-1}$ the scalar field mediates a fifth force that is $\alpha(\phi)$ times the strength of gravity.  We calculate $\alpha(\phi)$ by considering the conservation equation for $T_{\rm m}^{\mu \nu}$ which reads:
\be
\nabla_{\mu} T_{\rm m}^{\mu \nu} &=& \left[\beta(\phi) T_{\rm m} +\sum_{i}\beta_{i} S_{i}\right]\nabla^{\nu}\phi \\ &&- \beta(\phi)T_{\rm m}^{\mu \nu}\nabla_{\mu}\phi.\nonumber
\ee
For non-relativistic matter with energy density $\rho_{\rm m} = -T_{\rm m}^{0}{}_{0}$, $T_{\rm m} \approx -\rho_{\rm m}$, we find that a particle of matter, with mass $m$, feels an additional, or fifth, force $\vec{F}_{\phi}$ where:
\be\label{eq:scalarforce}
\vec{F}_{\phi} &=& -m\left[\beta(\phi) + \sum_{i} \beta_{i} \zeta_{i}\right] \vec{\nabla}\phi, \\
&=& -m\frac{\left[\beta(\phi) + \sum_{i} \beta_{i} \zeta_{i}\right]}{k(\phi)} \vec{\nabla}\varphi, \nonumber
\ee
where $\zeta_{i} = S_{i}/T_{\rm m} = -S_{i}/\rho_{\rm m}$.  When $\beta_{i}\zeta_{i}/\beta \ll 1$, it follows that, if $\phi$ is approximately massless (over the scale of interest), then it mediates a force that is $\alpha(\phi) = \beta^2 / k^2(\phi)$ times the strength of gravity, thus:
\be
\alpha(\phi) = \frac{\beta(\phi)^2}{\lambda_{}^{-2} + 3\beta^2(\phi)}.  \label{alphaEqn}
\ee
 Hence when $3\lambda^2_{} \beta^2 \gg 1$, the force is $1/3$ the strength of gravity (equivalent to an $\omega = 0$ Brans-Dicke theory), whereas in the opposite limit $\lambda^2_{} \beta^2 \ll 1$, we have $\alpha \approx \lambda^2_{} \beta^2(\phi) \ll 1/3$.

We estimated that if, today, $\phi$ lies near the minimum of the effective potential in the cosmological background, $\beta \approx 0.23$, and since we generally expect $\lambda$ to be $O(1)$ or larger, $\alpha \sim 0.05 - 1/3$.  We also noted that unless $\beta_{,\phi \phi}\gg 1$, $m_{\varphi} \sim O(H)$ and the force will be long range.    Solar system limits on long range fifth forces constrain $\alpha \lesssim 10^{-5}$ and so we must require that the theory possesses some mechanism so that, in the solar system either the new force becomes of short range (i.e $\lambda_{\varphi} = m_{\varphi}^{-1} \ll O(1)\,{\rm AU}$) , or the coupling, $\alpha(\phi)$, is suppressed.   The former scenario, where the range of the force is suppressed, has been dubbed the chameleon mechanism \cite{chameleon1,chameleon2,chameleon3} and relies heavily on the shape of the potential. We shall show that shape of $V(\phi)$ does not allow for a viable chameleon mechanism. On the other hand, the suppression of $\alpha(\phi)$ is the essence of the the Damour-Polyakov mechanism \cite{Damour:1994zq}. In Ref. \cite{Damour:1994zq}, whilst $\alpha(\phi)$ is minimized by the cosmological evolution of $\phi$, at any given epoch, $\alpha(\phi)$ does not exhibit very different values in different environments; hence if fifth forces are suppressed locally then they are also suppressed on cosmological scales.  In this article we propose a new mechanism inspired by that of Damour and Polyakov, so that although $\alpha(\phi)$ is suppressed in environments, such as the solar system, where the ambient density is much larger than the average cosmological density, cosmologically it is still possible for $\alpha(\phi) \sim O(1)$ leading to non-negligible effective modifications of  gravity on large scales. In the following sections we investigate these points further.

\section{Local Limits on Dilaton Models} \label{sec:Local}
\subsection{Local Constraints}
Whilst it seems natural that the scalar field(s) in dark energy theories should interact with ordinary matter, local tests of gravity tightly constrain any such coupling. If, in the solar system, a scalar field, $\varphi$, couples to ordinary matter with a strength $\alpha$ times that of gravity and has a mass, $m_{\varphi}$, then when $\alpha$ and $m_{\varphi}$ are both (at least approximately) constant, there is an additional force, $\vec{F}_{\varphi}$, between bodies with separation $r$.  When $r$ is much larger than the length scales of these bodies, $\vec{F}_{\varphi} = \alpha_{\rm eff}(m_{\varphi}r) \vec{F}_{N}$, where $\vec{F}_{N}$ is the usual Newtonian force, and $\alpha_{\rm eff} = \alpha(1+m_{\varphi}r)e^{-m_{\varphi}r}$.

When $m_{\varphi}r \ll 1$ and $\alpha_{\rm eff} \approx \alpha$, such a theory is approximately equivalent to Brans-Dicke theory with a Brans-Dicke parameter $\omega_{\rm BD} = 1/2\alpha_{\rm eff}-3/2$.  Current tracking of the Cassini satellite provides the best limit on this parameter: $\omega_{\rm BD} \gtrsim 40,000$ \cite{Bertotti:2003rm, Will:2001mx}. In this case $r$ was approximately the orbital radius of Saturn: $r \approx 9 - 10\,{\rm AU}$.  Thus:
\be
\alpha_{\rm local} < 1.2 \times 10^{-5}\quad {\rm if}, \quad m_{\varphi}^{{\rm local}-1}\gtrsim 9 - 10 \,{\rm AU}.
\ee

In Brans-Dicke models, matter is minimally coupled to a metric $\tilde{g}_{\mu\nu} = A^2(\phi)g_{\mu \nu}$.  If $\phi$ appears nowhere else in the matter action, this leads to a universal coupling between matter and $\phi$. More generally, and particularly with the dilaton, we expect $\phi$ will have additional couplings to matter. For instance, the gauge coupling `constants', $g_{i}$, are expected to depend on the value of dilaton field.   This leads to a non-universal coupling to matter which in turns violates the equivalence of free-fall i.e. the Weak Equivalence Principle (WEP).  There are tight laboratory constraints on any violation of WEP, which are parametrized in terms of $\eta_{\rm b-c}$:
\be
\eta_{\rm b-c} = \frac{2\vert \vec{a}_{b}-\vec{a}_{c}\vert}{\vert \vec{a}_{b}+\vec{a}_{c}\vert},
\ee
where $\vec{a}_{b}$ and $\vec{a}_{c}$ are respectively the accelerations of test mass B and test mass C towards some third body (which is generally the Sun).   Interpreting a given limit on $\eta_{\rm b-c}$ in terms of a limit on individual composition dependent couplings is not entirely straight-forward as there is some degeneracy between the different potential couplings, and the result depends on the composition of the test-masses.

In our model we have composition dependent couplings $\beta_{i}$ which respectively couple to a fraction $\zeta_{i}$ of a body's mass. In addition, there is the universal coupling $\beta$ which couples to all of the mass.   When $m_{\phi}r \ll 1$, the force between a body b and the Sun, is therefore a factor $\alpha_{\odot - {\rm b}}$ times that of gravity where:
\be
\alpha_{\odot - {\rm b}} = \frac{1}{k^2(\phi)}\sum_{i,j} \left[\beta + \beta_{i}\zeta_{\odot i}\right]\left[\beta + \beta_{j}\zeta_{b j}\right],
\ee
where $k^2 = 3\beta^2 +\lambda_{}^{-2}$. There is a similar expression for the force between a body c and the Sun, defining $\alpha_{\odot - {\rm c}}$.  We then have:
\be
\eta_{b-c} &=& \left\vert \alpha_{\odot-{\rm b}} - \alpha_{\odot -{\rm c}}\right\vert \nonumber \\&=&  \left\vert \frac{1}{3\beta^2 +\lambda_{}^{-2}}\sum_{i,j} \left(\beta + \beta_{i}\zeta_{\odot i}\right) \beta_{j}\zeta_{(b-c) j} \right \vert. \nonumber
\ee
where $\zeta_{(b-c) j} = \zeta_{b j}-\zeta_{c j}$.

The tightest current constraint on WEP violation was found by Schlamminger \etal \cite{Schlamminger:2007ht}, who measured the differential acceleration of two tests masses composed respectively of ${\rm Be}$ and ${\rm Ti}$ towards the Sun and found:
\be
\eta = (0.3 \pm 1.8) \times 10^{-13}.
\ee
In Ref. \cite{Dent:2008gu}, the dependence on the $\zeta_{i}$ was calculated. For simplicity we assume that only the fine structure constant $\alpha_{\rm em}$ and  $\Lambda_{\rm QCD}$ vary and that the lepton and quark masses are fixed.  We take $\beta_{\Lambda} = \delta \ln \Lambda/\delta \phi$ and $\beta_{\alpha} =\delta \ln \alpha/\delta \phi$.

For a substance  composed of atoms with baryon number $A$ and lepton number $Z$ we have:
\be
\zeta_{\alpha} \approx 2.7 \times 10^{-4} + (7.6 \times 10^{-4}) \frac{Z(Z-1)}{A^{4/3}}, \nonumber \\
\zeta_{\Lambda} \approx 0.95 + \left(1.7 \times 10^{-2}\right)A^{-1/3}. \nonumber
\ee
We approximate the composition of  the Sun as 75\% hydrogen and 25\% helium-4, and have:
\be
\zeta_{\alpha}^{\odot} \approx 3.0 \times 10^{-4}, \qquad \zeta_{\Lambda}^{\odot} \approx  0.97.   \nonumber
\ee
The  Schlamminger \etal  constraint \cite{Schlamminger:2007ht} gives:
\be
\left(-0.72\beta_{\alpha} + 1.7 \beta_{\Lambda}\right)\left(\beta + (3 \times 10^{-4})\beta_{\alpha} \right) \nonumber \\ \left. + 0.97 \beta_{\Lambda}\right)  < 10^{-10}(3\beta^2+\lambda_{}^{-2}). \nonumber
\ee
Now $\sqrt{\alpha} = \beta/k(\phi)$ and locally the Cassini limit gave $\beta/k < 3.5\times 10^{-3}$, and hence $3\beta^2 \ll \lambda_{}^{-2}$ and so $k \approx \lambda_{}^{-1}$ locally.  Assuming that $\beta/k$ is only just below this limit, the WEP violation constraint would require:
$$
-0.72\beta_{\alpha} + 1.7 \beta_{\Lambda} \lesssim  3\lambda_{}^{-1} \times 10^{-8},
$$
i.e. unless there is some exact cancellation $\beta_{i} \ll \beta$ locally.  In many cases, we would expect that $\beta \sim \beta_{\Lambda}$. However this is not necessarily the case, and, indeed, we shall see  that in our proposed model there would be no such link.  The limit from WEP violation apply when $m_{\phi}^{-1} \gtrsim 1\,{\rm AU}$.

Local limits on the matter to scalar field coupling are therefore particularly stringent, limiting it to be $\ll 1$ unless the scalar field is sufficiently heavy.  The Cassini and WEP violation bounds are evaded if the field has a Compton wavelength shorter than about ${\rm AU}^{-1}$, but a variety of different tests (for an excellent review see \cite{Will:2001mx}) require any gravitational strength fifth forces in the solar system to have a Compton wavelength of no more than about $0.1\,{\rm mm}^{-1}$.

\subsection{Chameleon Mechanism for the Dilaton}

Let us recall that dark energy models suffer from gravitational problems when coupled to ordinary matter. The Cassini bound is extremely stringent and implies that the dark energy scalar should be almost decoupled from matter. Of course, no known and well-motivated model of dark energy constructed so far satisfies this constraint. In general, couplings tend to be of order one (for couplings with much larger values and their phenomenology see \cite{chameleon3}). This would rule out most dark energy models. When the dark energy models are scalar-tensor theories with coupling to matter via the Jordan frame metric $\tilde g_{\mu\nu}= A^2 (\phi) g_{\mu\nu}$, the chameleon mechanism can alleviate these gravitational problems. This mechanism uses  two ingredients. The first one springs from the coupling of the dark energy field to matter and the resulting effective potential
\be
V_{\rm eff}(\phi; \rho_m)= \kap^2 (V(\phi) + \rho_m A(\phi))/2, \\
\square \phi = \frac{\kap^2}{2}V_{{\rm eff},\phi}(\phi,\rho_{\rm m}),
\ee
where $\rho$ is the pressure-less matter density. Notice that for a typical runaway potential $V(\phi)$, the matter contribution can induce a matter dependent minimum of the potential when $A(\phi)$ is an increasing function. The minimum $\phi_{\rm min}$ satisfies
\begin{equation}
V_{,\phi}(\phi_{\rm min})= -\rho_m A_{,\phi}(\phi_{\rm min})
\end{equation}
At this minimum, the field becomes massive with a mass
\be
m^2_\phi &=& V_{{\rm eff},\phi \phi}(\phi;\rho) =  \frac{\kappa_4^2}{2}\left[V_{\phi\phi}(\phi_{\rm min}) \right. \\ && \left.+\rho_m A_{,\phi\phi}(\phi_{\rm min})\right].\nonumber
\ee
Assuming that $V(\phi)$ and $A(\phi)$ are two convex functions, this mass is guaranteed to be bounded from below
\begin{equation}
m^2_\phi\ge  \frac{\kappa_4^2}{2} \rho_m A_{,\phi\phi}(\phi_{\rm min})
\end{equation}
which is a $\rho$- dependent quantity. As $\rho_m$ increases, the value of the minimum $\phi_{\rm min}$ decreases.
If the coupling function is such that $\rho_m A_{,\phi\phi}(\phi_{\rm min})$ increases with $\rho_m$, then the field $\phi$ may be massive enough to evade the gravitational tests in a dense environment. Typically, one requires that the mass of the scalar field should be larger than $10^{-3}$ eV.
This possibility can only be envisaged in relatively dense media such as the atmosphere. It would not be operative in a sparse environment such as the solar system vacuum. In this case, the scalar field may still be responsible for a distortion of planetary trajectories.
Fortunately, another mechanism can be at play: the thin shell effect.
Let us consider a spherical body and the influence that the scalar field generated by this body can have on particle trajectories.
Outside  the body, the force due to the scalar field is given by
\begin{equation}
\vec{F}_\phi= -\beta(\phi) m \vec{\nabla} \phi
\end{equation}
where $m$ is the mass of the test particle and $\beta$ is the value of the coupling where the force is evaluated (we have neglected composition dependent couplings, see eq. (\ref{eq:scalarforce})).
When the dark energy field is massless, the induced force is $\beta^2 \vec{F}_N$ where $\vec{F}_N$ is the Newtonian force. This leads to the Cassini constraints we have already mentioned. For a chameleon field with a density dependent mass, the effect is drastically different. Indeed, the field $\phi$ is mostly confined to vary within a thin shell close to the surface of the spherical body. The resulting force on a test particle is given by
\begin{equation}
\vec{F}_{\phi}= \frac{3\Delta R}{R}\beta^2  \vec{F}_N
\end{equation}
where we have assumed that $\alpha_\phi(\phi_m)$ is a very slowly varying function of $\phi_m$.  The size of the shell is given by
\begin{equation}
\frac{\Delta R}{R}=\frac{\vert \phi_\infty -\phi_c\vert}{ 3 \beta \Phi_N(R)}
\end{equation}
where $\phi_\infty$ is the value of the field at infinity and $\phi_c$ its value in the centre of the spherical body and we have defined the Newton potential at the surface of the body $\Phi_N(R)$ . We can see that the external force due to the scalar field is negligible provided the thin shell exists and therefore
\begin{equation}
{\vert \phi_\infty -\phi_c\vert}\ll { 3 \beta \Phi_N(R)}
\end{equation}
This criterion can be easily applied to  the case of dilatonic models.

Let us consider that the coupling function is an exponentially increasing function
\begin{equation}
A(\phi)= e^{\beta_0\phi}
\end{equation}
The normalised field $\varphi$ is nothing but a rescaling of $\phi$.
It is easy to see that the effective potential has a  minimum:
\begin{equation}
\phi_{\rm min}= \frac{1}{1-3\beta_0} \ln (\frac{\tilde V_0}{\rho}\frac{1-4\beta_0}{\beta_0})
\end{equation}
where $\beta_0\le 1/4$.
In a spherical situation, we have that
\begin{equation}
\vert \phi_\infty -\phi_c\vert= \frac{1}{1-3\beta_0}\vert \ln \frac{\rho_\infty}{\rho_c}\vert
\end{equation}
In the solar system, the density inside the sun does not exceed $\rho_c\sim 1\,{\rm g}\,{\rm cm^{-3}}$ while the vacuum is such that $\rho_\infty \sim 10^{-23} \,{\rm g}\,{\rm cm^{-3}}$.
Now the Newtonian potential at the surface of the sun is of order $\Phi_N (R)\sim 10^{-9}$ implying that a thin shell is not present. Hence this type of coupling for the dilaton does not lead to a chameleon mechanism and gravity would be greatly modified as $\alpha = \beta_0^2/k^2$ unless $\beta_0$ is  essentially zero to evade gravitational problems.

All in all, we have seen the thin shell mechanism does not exist for a dilaton when the coupling $\beta$ is nearly constant. Another way of satisfying the minimum equation is to compensate the large
variations of the matter density by an equivalently large variation of the coupling function $\beta$. In this case, gravity tests can be evaded using a different mechanism: the Damour-Polyakov effect whereby the coupling $\beta$ becomes density dependent and effectively vanishes in a dense environment.

\subsection{The Damour-Polyakov Mechanism}
Damour and Polyakov \cite{Damour:1994zq} proposed a mechanism whereby the coupling $\beta(\phi)$ would naturally be moved to small values by the expansion of the Universe.  The present smallness of $\beta(\phi)$ would then be a natural consequence of the age of the Universe.

The Damour-Polyakov mechanism assumes that $A(\phi)$ has a minimum at some $\phi = \phi_0$.    Hence near $\phi = \phi_0$ we have:
\be
A(\phi) \approx A_0\left[1 + \frac{A_2}{2}(\phi-\phi_0)^2\right].
\ee
We may always rescale $A$ and the metric to fix $A_0 = 1$.  Near the minimum where $A_{2}(\phi-\phi_0)^2/2 \ll 1$, we then have $\beta(\phi) = (\ln A)_{,\phi} \approx A_{2}(\phi-\phi_0)$.  Thus the field is sufficiently near the minimum $\alpha = \beta(\phi)^2/k^2(\phi) \approx \lambda^2 A_2^2(\phi-\phi_0)^2 \ll 1$ and so fifth force effects are suppressed.  To suppress WEP violation effects one would also have to ensure that the $\beta_{i}$ are $\ll 1$ near $\phi = \phi_0$.  The original model assumes negligible potential, but it would continue to work well provided that $V(\phi)$ is also minimized at $\phi_0$.  In either case, whatever value $A_2$ takes, the cosmological evolution of $\phi$ will drive $\phi$ towards $\phi_0$.  If, however,   as in our case we have a potential $V(\phi) = A^4(\phi) V_0 e^{-\phi}$ which represents dark energy today, $V(\phi)$ is then non-negligible and the exponential part of this potential means that $V(\phi)$  is \emph{not} minimized at $\phi=\phi_0$.

Let us assume that at early times the matter coupling dominates the evolution of $\phi$ and $\phi$ is driven close to $\phi_0$, with $\vert \phi -\phi_0\vert \ll 1$.  Then
\be
V &\approx& V_c\left[1-(\phi-\phi_0) + \frac{1+4A_2}{2}(\phi-\phi_0)^2 \right. \\  \nonumber
&& \left.+ O\left(\phi-\phi_0)^3\right)\right], \nonumber
\ee
where $V_c = V_0 \exp(-\phi_0) $. Requiring that $V(\phi)$ drives the present period of acceleration expansion we have $\kap^2 V_c \sim   3\Omega_{\Lambda 0}H^2_0$.

Now since $\vert \phi-\phi_0\vert \ll 1$ we have that $\alpha = \beta^2(\phi)/k^2(\phi) = 1/(3 + 1/\lambda^2 \beta^2)$ and to expect $\alpha \ll 1$ at late times, we must have $3\lambda^2 \beta^2 \ll 1$ so $k(\phi) \approx \lambda^{-1}$.  It follows from $\dd \varphi = k(\phi) \dd \phi$ that $\varphi = \phi/\lambda$.

Defining  $\delta \phi = \phi-\phi_0$ and assuming $A_2 \delta \phi^2/2 \ll 1$, the field equation for $\delta \phi$  becomes:
\be
-\delta\ddot{\phi} - 3H\delta\dot{\phi} &\approx& -\frac{\lambda^2 \kap^2}{2}V_c e^{-\delta\phi} \label{cos1} \\ &&+ \frac{\lambda^2 \kap^2}{2} \left[\rho_{\rm m}+4V_c e^{-\delta\phi}\right] A_2 \delta \phi.  \nonumber
\ee
Now $\beta(\phi) = A_2 \delta \phi$ and for consistency with local tests we must require $\vert \beta \vert \ll 1$.   Therefore, assume that at the end of the matter era, the cosmic evolution has ensured  $\vert \beta \vert \ll 1$.  In the current epoch $V(\phi) \approx V_c e^{-\phi} \sim O(\rho_{\rm m})$, so the second term on the right hand side of Eq. (\ref{cos1}) is roughly a factor $\beta$ smaller than the first, and hence initially $\delta \phi$ would evolve according to:
\be
-\delta\ddot{\phi} - 3H\delta\dot{\phi} &\approx& -\frac{\lambda^2 \kap^2}{2} V_c e^{-\delta\phi}.
\ee
If $\beta(\phi)$ remains $\ll 1$ small then since $\kap^2 V_c/2 \sim O(H^2)$ today, $\delta \phi$ would move by roughly $O(\lambda^2)$, which is expected to be $O(1)$ or greater, so finally we expect $\vert \delta \phi \vert \gtrsim O(1)$.  For $\vert\beta\vert =  \vert A_2 \delta \phi\vert \ll 1$ to remain the case we would have to require $A_2 \ll 1$.    With $A_2$ small enough to satisfy constraints from fifth-force tests, the resulting theory would, essentially, evolve just like an uncoupled quintessence field in an exponential potential; we would also have to enforce that $\lambda^2$ is small enough that the equation of state of this quintessence is approximately $-1$ as is observed.

The condition $A_2 \ll 1$ followed directly from the assumption that $\vert \beta(\phi) \vert \ll 1$ cosmologically at the present time for consistency with local tests.  However, local tests only imply a small value of $\vert \beta\vert$ locally i.e. in the solar system. If the local and cosmological values of $\beta$ are not equal, we would not necessarily  need to require that $\vert \beta \vert \ll 1$ and hence $A_2 \ll 1$ cosmologically today.  Additionally if $\beta \sim O(1)$ cosmologically whilst $\beta \ll 1$ locally, we would have interesting modifications to gravity emerging on large scales.    We now show that such a mechanism exists and is compatible with local tests if $A_2 \gg 1$.  This is essentially an environmental dependent version of the standard Damour-Polyakov mechanism, where the coupling is minimized by fact that the local matter density is much greater than the cosmological one.

\section{Environmentally Dependent Dilaton}\label{sec:Environ}
In this section we propose a dilaton theory whereby the dilaton acts as the
dark energy particle and is both consistent with local tests but leads to non-negligible deviations from General Relativity on astrophysical scales.  In common with chameleons theories, the properties of the dilaton field exhibit an environmental dependence which ensure that, in high density environments, dilaton mediated fifth-forces are negligible whilst allowing them to be of gravitational strength  in low density regions.  In chameleon models, the strength of the scalar-field to matter coupling is roughly constant, but the mass grows with the ambient density. This leads to a very short range fifth force in the laboratory but a relatively long range force in the cosmological background.    In our environmentally dependent dilaton model, however, the scalar is generally light locally (i.e. there is a long range fifth force), but the matter coupling itself is minimized in high density regimes.  This may be seen as an environmentally dependent analogue of the Damour-Polyakov mechanism.  Instead of the coupling tending to zero at late-times, the coupling decreases as the density increases.  We therefore refer to the mechanism by which the coupling is suppressed locally as the Environmentally Dependent Damour-Polyakov (EDDP) mechanism.

\subsection{The model}
We consider a dilaton model described, in the Einstein frame by the action given in Eq. (\ref{action}).  In common with the standard Damour-Polyakov mechanism we assume that:
$$
A(\phi) \approx 1 + A_2(\phi-\phi_0)^2/2,
$$
when $A_2(\phi-\phi_0)^2/2 \ll 1$.  We check that this quantity is indeed small later.    It follows that:
\be
\beta(\phi) &\approx& A_2 (\phi-\phi_0), \\
k(\phi) &=& \sqrt{3A_2^2(\phi -\phi_0)^2 + \lambda^{-2}}.
\ee
We take $A_2 \gg 1$.  We shall show below that $A_2 \gg 1$ is actually required for the coupling to be sufficiently suppressed locally as to evade fifth-force constraints.

The largeness of the required value of $A_2$ certainly appears unnatural on a first inspection.  It is, however, important to note that there is already a large parameter in this theory namely $c_1 = e^{-\psi(\phi_0)} \equiv  M_{\rm pl}/M_{\rm s}$.  It is feasible that the largeness of $A_2$ is linked to the discrepancy between the string and 4-d Planck scales. For instance, let us define $\chi = M_{\rm pl}\phi$, so that $\chi$ has the canonical units of mass. Then:
\be
A \approx 1 + \frac{A_2}{2M_{\rm pl}^2}(\chi-\chi_0)^2 = 1+\frac{A_2}{2c_1^2}\left(\frac{\chi-\chi_0}{M_{s}}\right)^2. \nonumber
\ee
If the minimum in $A$ at $\phi = \phi_0$ ($\chi = \chi_0$) is due to some non-perturbative effect associated with the string mass scale, $M_{\rm s}$, one would actually expect the coefficient of $((\chi-\chi_0)/M_{\rm s})^2$ to be $O(1)$, and so $A_2 \sim O(c_1^2) \gg 1$. Thus, although  demanding $A_2 \gg 1$ is certainly not as aesthetically pleasing a requirement as $A_2 \sim O(1)$, it is not necessarily unnatural if it is associated with the fact that $M_{\rm s} \ll M_{\rm pl}$.

The effective potential, $V_{\rm eff}$, is given by $V_{\rm eff} = V_0 A^4(\phi)e^{-\phi} - A(\phi)T_{\rm m}$.  Thus in  a background where the matter is non-relativistic so $T_{\rm m} \approx -\rho_{\rm m}$,  this effective potential is minimized when $\beta(\phi)\left[\rho_{\rm m} + 4 A^3(\phi)V_0 e^{-\phi}\right] = A^3(\phi) V_0 e^{-\phi}$.  Assuming $A(\phi) \approx 1$, this is achieved when:
\be
\beta(\phi_{\rm min}) &=& A_2(\phi_{\rm min} -\phi_0) \\ &\approx& \frac{V_{0}e^{-\phi_{\rm min}}}{\rho_{\rm m} + 4 V_{0}e^{-\phi_{\rm min}}} \leq 1/4. \nonumber
\ee
We note that as $\rho_{\rm m} \rightarrow \infty$, $\beta(\phi_{\rm min}) \rightarrow 0$.

Now if $A_2 \gg 1$,  $\phi_{\rm min} -\phi \ll 1/4A_2 \ll 1$ and so we may replace $V_0 e^{-\phi_{\rm min}}$ by $V_0 e^{-\phi_0} = V_c$.   Additionally:
$$
\frac{A_2}{2}(\phi_{\rm min}-\phi_0)^2 \leq  \frac{1}{32A_2} \ll 1,
$$
for all $\rho$.  The assumption that $A_2(\phi-\phi_0)^2/2 \ll 1$ and $A(\phi) \approx 1$ is therefore valid because $A_2 \gg 1$.

Finally, with $A_2 \gg 1$, we have that the dominant contribution to the mass, $m_{\varphi}$, of small perturbations in the scalar field about its effective minimum , $\phi_{\rm min}(\rho)$, in a background with density $\rho_{\rm m}$ and $\phi = \phi_{\rm min}$ is given by:
\be
m_{\varphi}^2(\phi_{\rm min}) \approx \frac{\kap^2 A_{2}}{2k^2(\phi_{\rm min})}\left[\rho_{\rm m} +4V_c\right],
\ee
where $k^2(\phi_{\rm min}) = 3\beta^2(\phi_{\rm min}) + \lambda^{-2} < \lambda^{-2} + 3/16$.  In a background where $\phi = \phi_{b}$, and over scales $r \ll 1/m_{\varphi}$, the fifth force mediated by $\phi$ between two point masses is  $\alpha_{\phi}$ times the strength of the particles' mutual gravitational attraction, where:
\be
\alpha_{\phi}(\phi_b) = \frac{\beta^2(\phi_b)}{k^2(\phi_b)}  = \frac{\beta^2(\phi_b)}{3\beta^2(\phi_b)+\lambda_{}^{-2}}.  \nonumber
\ee
Hence when $\sqrt{3}\beta(\phi_b) \lambda^2 \gg 1$, $\alpha_{\phi} \approx 1/3$ and in the opposite limit $\alpha_{\phi} \approx \lambda_{}^2 \beta^2$ which is minimized as $\beta \rightarrow 0$ i.e. $\phi \rightarrow \phi_0$.    We note that  the value of $\beta$ at the minimum of the effective potential tends to zero as the ambient density of matter, $\rho_{\rm m}$, grows large: $ \beta \propto 1/\rho_{\rm  m}$.  Thus as $\rho_{\rm m} \rightarrow \infty$, $\alpha_{\phi} \propto 1/\rho_{\rm m}^2 \rightarrow 0$.

With $\vert\phi-\phi_{\rm min}\vert \ll 1$ and  $A(\phi) \approx 1$, $\beta(\phi)\approx A_2(\phi-\phi_0)$ and $V(\phi) = V_0 e^{-\phi} \approx V_{0}e^{-\phi_{\rm min}} = V_{c}$, the field equation for $\phi$ in a general background is then:
\be
\square \varphi = \frac{\kap^2}{2k(\phi)}\left[-V_c + \beta(\phi)(4V_c -T_{\rm m})\right], \label{phi:eq}
\ee
where $\dd\varphi = k(\phi) \dd \phi$, $k(\phi) = \sqrt{3\beta^2(\phi) + \lambda_{\rm s}^{-2}}$.

\subsection{Background Cosmological Behaviour}
We now consider the large scale, homogeneous and isotropic cosmological behaviour of $\phi$ in this model.   We take the background metric to be that of FRW spacetime with dust matter so $T_{\rm m} = - \rho_{\rm m}$:
\be
\dd s^2 = -\dd t^2 + a^2(t)\left[ \dd r^2 + f_{k}(r)^2\dd \Omega^2 \right],
\ee
where $f_{k} = \sin(\sqrt{k}r)/\sqrt{k}$.  We define $\Omega_{\rm m}$ and $\Omega_{\Lambda}$ by $\kap^2 \rho_{\rm m} = 3\Omega_{\rm m}H^2$ and $\kap^2 V_{c} = 3\Omega_{\Lambda}H^2$.  With these definitions the value of $\phi$ at the minimum of the effective potential is $\phi_{\rm min}$, where:
\be
\beta(\phi_{\rm min}(t)) = A_{2}(\phi_{\rm min}(t)-\phi_0) = \frac{\Omega_{\Lambda}(t)}{\Omega_{\rm m}(t) + 4\Omega_{\Lambda}(t)}.
\ee
The parameters are such that the dilaton is consistent with being dark energy.
We take $A_{2}\gg 1$  (which we shall see is required by local test below).  The field equation for $\phi$ in this background becomes:
\be
-\ddot{\varphi} - 3H \dot{\varphi} = -\frac{\kap^2 V_{c}}{2k(\phi)} +  \frac{A_{2}(\phi-\phi_{\rm 0})}{k(\phi)}  \left[\rho_{\rm m} + 4V_c\right].
\ee
This is solved approximately by:
\be
\phi \approx \phi_{\rm min}(t) = \frac{V_{c}}{\rho_{\rm m}(t)+4V_c}. \nonumber
\ee
This is the valid leading order approximation to the solution provided $m_{\varphi}^2 \gg H^2$, where $m_{\varphi}$ is the mass of small perturbations in $\varphi$ about the minimum of the effective potential. We have:
\be
\frac{m_{\varphi}^2}{H^2} \approx \frac{3A_{2}}{2} (\Omega_{\rm m} + 4\Omega_{\Lambda}) \left[\lambda_{}^{-2} + 3\left(\frac{\Omega_{\rm m}}{\Omega_{\Lambda}} +4\right)^{-2}\right]^{-1}. \label{eq:cosmomass}
\ee
With $\lambda_{} \sim O(1)$ or greater, and $\Omega_{\rm m} + \Omega_{\Lambda} \sim O(1)$ it is clear that typically $m_{\varphi}^2/H^2 \sim O(A_2)$ or larger.  Since $A_2 \gg 1$, it follows that $m_{\varphi} \gg H$ as required.    Thus $\beta(\phi) \approx \beta(\phi_{\rm min})$ cosmologically.  Since $\beta(\phi_{\rm min}) \leq 1/4$, it follows that $\phi - \phi_{0} < 1/4A_2 \ll 1$, justifying the assumption that $\vert \phi-\phi_{0} \vert \ll 1$.

Additionally, since $m_{\varphi}^2 \gg H^2$, the background cosmology is observationally indistinguishable from that of a  $\Lambda$CDM model with $\Lambda = \kap^2 V_c$.  This said, provided $A_2$ is not too large, we shall see in \S \ref{sec:Struct}, that the force mediated by $\phi$ is still sufficiently long range to have a detectable effect on the formation of large scale structures.  For linear perturbations in the matter density and over spatial scales smaller than $m_{\varphi}^{-1}$, this new force is $\alpha_{\rm cos}$ times that of gravity where:
\be
\alpha_{\rm cos} = \frac{\beta^2(\phi_{\rm min})}{k^2(\phi_{\rm min})} = \frac{1}{3+ \lambda_{}^{-2}(4+\Omega_{\rm m}/\Omega_{\Lambda})^2}.
\ee
Taking $\Omega_{\rm m} = 0.27$ and $\Omega_{\Lambda} =0.73$, we have $\Omega_{\rm m}/\Omega_{\Lambda} \approx 0.37$ today, and $\lambda_{}\gtrsim 1$, we have
 $0.045 \lesssim \alpha_{\rm cos} < 1/3$ today. The larger values of $\alpha_{\rm cos}$ correspond to the larger values of $\lambda_{}$.  Thus $\alpha_{\rm cos}$ is much larger than the Cassini limit on the local value of $\alpha$ (i.e. $10^{-5}$).

We note that we have $\alpha_{\rm cos} \approx 1/3$ if $\lambda_{} \gg 4.4$. The cosmological value of $\beta$ today, $\beta_{\rm cos}$, is:
\be
\beta_{\rm cos} = \frac{1}{4+\Omega_{\rm m}/\Omega_{\Lambda}} \approx 0.23.
\ee
Therefore, on scales much smaller than the horizon but potentially cosmologically still important (depending on the size of $A_2$), the scalar field mediates a force with a strength slightly smaller than gravity. This will influence structure formation on those scales.

\subsection{Local Behaviour and Constraints}
Previously we found that $\alpha_{\rm \phi}(\phi_{\rm min}) \rightarrow 0$ as $\rho_{\rm m} \rightarrow \infty$.   We now show that, provided $A_2$ is large enough, the presence of our galaxy is enough to reduce $\alpha_{\rm phi}$ from its cosmological value, $\alpha_{\rm cos} \sim O(1)$, to a value locally, $\alpha_{\rm local}$, that is small enough to evade the Cassini limit and other local tests ($\alpha_{\rm local} < 10^{-5}$).

Firstly, since $\dd \varphi = k(\phi) \dd \phi = k(\phi) \dd \beta / A_2$, we can rewrite  Eq. \eqref{phi:eq} (with $T_{\rm m} = -\rho_{\rm m}$) in terms of $\beta$:
\be
\square \beta + \frac{3\beta}{3\beta^2 + \lambda_{}^{-2}} (\nabla \beta)^2 &=& \frac{4\pi G_N A_2}{3\beta^2 + \lambda_{}^{-2}}\left[-V_c \right. \\ && \left.+ \beta \left(\rho_{\rm m} + 4V_c\right)\right]. \nonumber
\ee
Defining $X = 3\beta^2/2 + \lambda_{}^{-2}\ln \beta$, so that $\dd X = (3\beta  + \lambda_{}^{-2}/\beta)\dd \beta$, one can rewrite the last equation further:
\be
\square X + \lambda_{}^{-2}(\nabla \ln \beta)^2 = \\
4\pi G_N A_2\left[-\frac{V_{c}}{\beta} + \left(\rho_{\rm m} + 4V_c\right)\right]. \nonumber
\ee

\paragraph{Effect of a local over-density:}: Consider the perturbation in $X$, $\delta X$, created by a quasi-static sub-horizon perturbation in $\rho_{\rm m}$, $\delta \rho_{\rm m}$.  The quasi-static and sub-horizon conditions imply that the time scale over which $\delta \rho_{\rm m}$ evolves is much longer than the typical length scale of the perturbation.  It also means that we can take $\square X \rightarrow \vec{\nabla}^2 \delta X$.    Far from the perturbation we take $\phi \rightarrow \phi_{\infty}$, $\rho \rightarrow \rho_{\infty}$, $\beta \rightarrow \beta_{\infty}  = A_2(\phi_{\infty}-\phi_0)$.

We set  $\rho_{\rm m} = \rho_{\infty} + \delta \rho_{\rm m}$, so that far from the perturbation, $\delta \rho_{\rm m} \rightarrow 0$, and $X \rightarrow X_{\infty} = 3\beta_{\infty}^2/2 + \lambda^{-2}_{} \ln \beta_{\infty}$. We define the shorthand notation $\beta(r) = \beta(\phi(r))$. We also assume that far from the perturbation, $\phi$ lies close to the minimum of its effective potential so that:
\be
\beta_{\infty} = \frac{V_{c}}{\rho_{\infty} + 4V_c}. \nonumber
\ee
Writing $\beta = \beta_{\infty} + \delta \beta$, $X=X_{\infty} + \delta X$, and \emph{without} linearizing, the field equation becomes:
\be
\vec{\nabla}^2 \delta X &=& 4\pi G_{N} A_2 \delta \rho_{\rm m} + 4\pi G_{N} A_2 V_{c} \left(\frac{\delta \beta}{\beta_{\infty} \beta}\right) \\
&& - \lambda_{}^{-2}(\vec{\nabla} \ln \beta)^2,  \nonumber
\ee

We now simplify to the case where $\delta \rho_{\rm m}$ represents a spherically symmetric over-density of matter, which is non-decreasing as $r\rightarrow 0$ i.e $\dd \delta \rho_{\rm m}(r)/ \dd r \leq 0$ and hence $\delta \rho_{\rm m}(r) \geq 0$.  The effect of the perturbation is to move $X$ and hence $\beta$ to smaller values i.e. $\beta(r) \leq \beta_{\infty}$ and more over from $\dd \delta \rho_{\rm m}(r)/ \dd r \leq 0$, $\dd \beta / \dd r \geq 0$. We now construct a necessary condition for $\beta(r)$ to be smaller than some $\beta_{\ast}$ for $\delta \rho_{\rm m} \geq 0$.

From $\beta \leq \beta_{\infty}$ we have:
$$(3\beta_{\infty}^2+\lambda_{}^{-2}) (\delta \beta)/\beta \leq \delta X  \leq 0$$
and so:
\be
\frac{4\pi G_{N} A_2 V_{c}}{\beta_{\infty}} \frac{\delta \beta}{\beta} &\leq& \frac{4\pi G_{N} A_2 \left[\rho_{\infty} + 4V_{c}\right]}{3\beta_{\infty}^2+\lambda_{}^{-2}} \delta X \\ && \equiv m^2_{\infty} \delta X \leq 0, \nonumber
\ee
where the last line is definition of $m_{\infty}^2$.  It is clear that $- \lambda_{}^{-2}(\vec{\nabla} \ln \beta)^2 \leq 0$, and so we have: $\delta \bar{X} < \delta X \leq 0$ where:
\be
\vec{\nabla}^2 \delta \bar{X} = 4\pi G_{N} A_2 \delta \rho_{\rm m}. \nonumber
\ee
The perturbation in the Newtonian potential, $\delta \Phi_{N}$, due to $\delta \rho_{\rm m}$ is given by:
$$
\vec{\nabla}^2 \delta \Phi_{N} = 4\pi G_{N}\delta \rho_{\rm m},
$$
where $\delta \Phi_{N} \rightarrow 0$ as $r \rightarrow \infty$. Hence we have $\delta \bar{X} = A_2 \delta \Phi_{N}$.
$\beta(r) < \beta_{\ast}$  is equivalent to $X(r) < X_{\ast} = 3\beta_{\ast}^2 + \lambda_{}^{-2} \ln \beta_{\ast}$. It follows that a necessary condition for $\beta(r) < \beta_{\ast}$ is that $\delta \bar{X} = A_2 \delta \Phi_{N} < X_{\ast}-X_{\infty}$:
\be
3(\beta_{\infty}^2-\beta_{\ast}^2) + \lambda_{}^{-2} \ln\left(\frac{\beta_{\infty}^2}{\beta_{\ast}^2}\right) < 2A_2 \vert \delta \Phi_{N}(r)\vert. \label{ness:cond}
\ee
In such a set-up, we also have
\be
\beta(r) \geq \beta_{\rm min}(r) = \frac{V_{c}}{\rho_{\infty} +\delta \rho(r) + 4V_{c}}. \nonumber
\ee
Hence for $\beta(r) < \beta_{\ast}$ it is also necessary that:
\be
\beta_{\rm min}(r) = \frac{V_{c}}{\rho_{\infty} +\delta \rho(r) + 4V_{c}} < \beta_{\ast}. \label{ness:cond2}
\ee
Eqs. \eqref{ness:cond} and \eqref{ness:cond2} are only necessary but \emph{not} sufficient to ensure that $\beta(r) < \beta_{\ast}$.  Given a limit on $\alpha_{\rm local}$, it does, however, allow us to place a lower bound on $A_2$ given $\lambda_{}$.   It should be noted that although Eqs. \eqref{ness:cond} and \eqref{ness:cond2} are only necessary conditions, when $\dd \rho / \dd r \leq 0$,  the combination of the two conditions is often almost sufficient i.e. if $\delta \bar{X} \ll X_{\ast}-X_{\infty}$ and $\beta_{\rm min}(r) \ll \beta_{\ast}$, then typically $\beta(r) < \beta_{\ast}$.

\paragraph{Application to Our Galaxy:} Laboratory, satellite and other solar system (collectively `local') tests of gravity certainly take place inside a sizeable over-dense region, namely our galaxy. The matter in our galaxy has the effect of moving $\beta$ to a smaller value locally than it has cosmologically.   There will also be additional contributions to the local value of $\delta \beta$ coming from the other members of the local group and local super cluster.  For the purpose of the paper though we take a cautious approach and are only concerned with how big $A_2$ must be so that the presence of the (roughly spherically) dark matter halo of galaxy is, alone, enough to ensure compatibility with local tests.  Specifically we focus on the Cassini limit $\alpha_{\rm local} < 10^{-5}$.

\begin{figure}[tbh]
\begin{center}
\includegraphics[width=7.5cm]{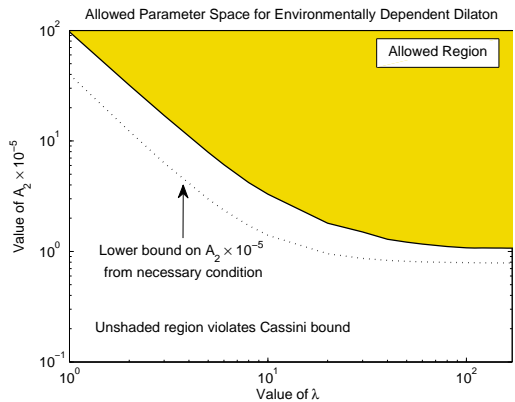}
\caption[]{Allowed parameter space for the environmentally dependent dilaton model.  The shaded region is that where the presence of our galaxy is sufficient to ensure that the local value of the fifth force coupling, $\alpha$, is smaller than the Cassini probe upperbound of $10^{-5}$.   We have modelled the galaxy as a spherical dark matter halo with NFW profile.  We have taken typical values for the NFW model parameters for our galaxy: $r_{\rm vir} = 267\,{\rm kpc}$,  $c=12.0$, $M_{\rm v} = 0.91\times 10^{12}M_{\odot}$.  We take the galactocentric radius of the solar system, $r_{\odot}$ to be $r_{\odot} \approx 8.3\,{\rm kpc}$. These choices correspond to $\Phi(r_{\odot}) = 1.02\times 10^{-6}$ and $\rho(r_{\odot}) = 0.22 {\rm GeV}\,{\rm cm}^{-3}$.  This value for $\rho(r_{\odot})$ limits $\lambda_{} < 170$, and we have plotted the constraints on $A_{2}$ for $\lambda_{} \in [1,170]$.   Very similar bounds on $A_{2}$ result for different realistic models of the galactic halo.
\label{fig:bounds}}
\end{center}
\end{figure}

We consider the idealized situation where our galaxy's dark matter halo is a spherically symmetric overdensity $\delta\rho = \rho_{\rm gal}(r)$ where $\dd \rho_{\rm gal} / \dd r \leq 0$.  Assuming that our galaxy sits in a region of cosmological density, $\rho_{\infty} = \rho_{\rm cos}^{\rm today} = 3\Omega_{\rm m 0}H_{0}^2/2\kap^2$, and hence $\beta_{\infty} = \beta_{\rm cos}^{\rm today} \approx 0.23$.  We assume that all local tests take place at a galactocentric radius $r_{\odot}$, so $\alpha_{\rm local} = \alpha(\phi(r_{\odot}))$.

We take $\delta \rho_{\rm gal}$ to be described by an NFW profile with core radius $r_{\rm c}$ and virial mass $M_{\rm v}$:
\be
\kap^2 \delta \rho_{\rm m}(r) = \frac{2GM_{\rm v}g(c)}{r_{c}^3}\frac{1}{x(1+x)^2},
\ee
where $x = r/r_{c}$; $g(c) = \left[\ln(1+c)-c/(1+c)\right]^{-1}$. Here, the virial mass is defined as being the mass inside $r_{500}$ which itself is defined as the radius inside which the average density is 500 times that of the cosmological background; the concentration parameter, $c$, is $r_{500}/r_{\rm c}$.

From this we have:
\be
\delta \Phi_{\rm N}(r) = - \frac{GM_{v}g(c)}{r}\ln\left(1+\frac{r}{r_{c}}\right).
\ee
Now $\alpha_{\rm local} = \beta^2(r_{\odot})/(3\beta^2(r_{\odot})+\lambda^{-2})$ and so $\alpha_{\rm local} \ll 1$ requires $\beta^2(r_{\odot}) \ll \lambda^{-2}/3$ in which limit $\alpha_{\rm local} \approx \beta^2(r_{\odot})/\lambda^2$.  Thus, using Eq. \eqref{ness:cond}, for $\alpha_{\rm local} < 10^{-5}$  it is necessary that:
\be
A_2 > 10^{6} \left[ 0.08 + 4\lambda^{-2}_{\rm s} +0.5\lambda^{-2}_{\rm s}\ln \lambda^2\right]\left \vert \frac{1.0 \times 10^{-6}}{\delta \Phi_{N}(r_{\odot})}\right \vert.
\ee
Typically values of  $M_v$, $c$, $r_s$ and $r_{\odot}$  from fitting measured rotation curves to an NFW profile \cite{Xue:2008se,Amsler:2008zzb} are:
\be
&M_{v} = 0.91^{+0.27}_{-0.18} \times 10^{-12} M_{\odot}, \qquad  &c = 12.0 \pm 0.3, \nonumber\\
&r_{\rm vir} = 267^{24}_{-19}\,{\rm kpc},  \qquad  &r_{\odot} = 8.0 \pm 0.5\,{\rm kpc} \nonumber
\ee
which give:
\be
\delta \Phi_{\rm N}(r_{\odot}) \approx 0.75 -1.4 \times 10^{-6}, \nonumber \\
\rho_{\rm gal}(r_{\odot}) \approx 0.15 - 0.37\, {\rm GeV}\,{\rm cm}^{-3}.\nonumber
\ee
Other estimates give $\rho_{\rm gal}(r_{\odot}) \approx 0.1 - 0.7\, {\rm GeV}\,{\rm cm}^{-3}$.  Putting this value into Eq \eqref{ness:cond2}, and requiring $\alpha_{\rm local} < 10^{-5}$ then gives:
\be
\lambda_{} \lesssim (0.77 - 5.4)\times 10^{2}.
\ee

We find more accurate limits on $A_2$ by numerically integrating the field equations for $\phi$ with an NFW matter profile. The constraints dependent on $r_{\odot}/r_{c}$ and $GM_{\rm v}g(c)/r_{\rm c}$.  We take typical values $r_{\odot}/r_{c} = 0.37$ (corresponding to $r_{\odot} = 8.3\,{\rm kpc}$, $r_{\rm vir} = 267\,{\rm kpc}$ and $c=12.0$), and $GM_{\rm v}g(c)/r_{\rm c} = 1.2 \times 10^{-6}$ (corresponding to $M_{\rm v} = 0.91\times 10^{12}M_{\odot}$); this gives $\Phi(r_{\odot}) = 1.02\times 10^{-6}$ and $\rho(r_{\odot}) = 0.22 {\rm GeV}\,{\rm cm}^{-3}$.  This value for $\rho(r_{\odot})$ limits $\lambda < 170$.  For these values of the parameters, the constraints on $A_2$ for $\lambda \in [1,170]$  for these are plotted in FIG. \ref{fig:bounds}. For $\lambda_{} \gg 1$, the analytic necessary (but not sufficient) lower bound on $A_{2}$ is only a factor $\approx 1.3$ smaller than the necessary and sufficient lower bound found by numerically integrating the equations. For $O(1)$ values of $\lambda$, the discrepancy increases to roughly a factor of $2$.  Nonetheless, it is clear that the lower limit on $A_2$ from the necessary condition of Eq. (\ref{ness:cond}) is generally within a factor of a few of the necessary and sufficient limit on $A_2$.  Whilst the upper limit on $\lambda$ is sensitive to the local halo density, very similar limits on $A_{2}$ for given $\lambda_{}$  are found for different realistic values of the NFW parameters.

\begin{figure}[tbh]
\begin{center}
\includegraphics[width=7.5cm]{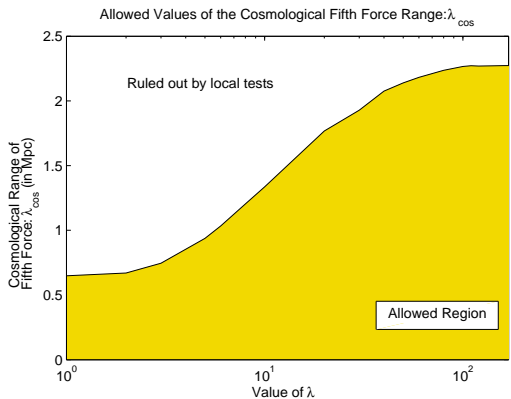}
\caption[]{Allowed values of the cosmological force range, $\lambda_{\rm cos}$, today given compatibility with the Cassini constraint on $\alpha_{\rm local}$.  Allowed values of $\lambda$ must be $\lesssim 170$ and are expected to be $\gtrsim 1$ and so we show only the region $\lambda \in [1,170]$.  We see that $\lambda_{\rm cos} \lesssim 0.5 - 2.2 {\rm Mpc}$ today.
\label{fig:lambda}}
\end{center}
\end{figure}

We briefly comment on limits from WEP violation.  We found in \S \ref{sec:Local} that if $\alpha_{\rm local}$ is only just below the Cassini limit of $\approx 10^{-5}$ then the WEP violating couplings $\beta_{\alpha}$ and $\beta_{\Lambda}$ would have to be less than $3 \times 10^{-8}\lambda_{}^{-1}$.  In our model $\beta_{i} \propto \tilde{V}(\phi) \propto \exp(-\phi)$.  In such a model $\tilde{V}(\phi) \sim M_{\ast}^4 e^{-\phi}$ where $M_{\ast}$ is typically of the same order of magnitude as the other mass-scale in the high energy theory.  Today $\tilde{V}(\phi) \approx M_{\Lambda}^4 \ll M_{\ast}^4$ (where $M_{\Lambda} = 2.4 \pm 0.3 \times 10^{-3}\,{\rm eV}$). This suppression is achieved by having $\phi \gg 1$ so $\exp(-\phi) \ll 1$.   Thus the same mechanism that suppresses the mass-scale of the effective cosmological constant would suppress the WEP violating couplings and naturally result in them having small value. For instance $\beta_{i} \lesssim 10^{-10}$ would only imply $\exp(-\phi) \sim 10^{-10}$ and so $M_{\ast} \gtrsim 0.76\,{\rm eV}$.  If $M_{\ast} \sim \Lambda_{\rm QCD} \sim 100\,{\rm MeV}$, say, then we would expect $\beta_{i} \sim \exp(-\phi) \sim 10^{-43}$.  It is therefore natural for WEP violation to be suppressed in this model.

In the previous subsection, we found (see Eq. \eqref{eq:cosmomass}) that the cosmological mass-squared, $m_{\rm \varphi}^2$, of small perturbations in the dilaton field was proportional to $A_{\rm 2}$, as well as having some dependence on $\lambda_{}$.  We can now transform our lower bounds on $A_{2}$ to upper bounds on the range, $\lambda_{\rm cos} = \hbar c/ m_{\varphi}^{\rm (cos)}$, of the dilaton mediated fifth force in the cosmological background.    The allowed values of $\lambda_{\rm cos}$ today (in units of Mpc) are shown in  FIG. (\ref{fig:lambda}).  We see that for values of $\lambda_{}$ and $A_{2}$ allowed by local tests, $\lambda_{\rm cos} \lesssim 0.5 - 2.2 {\rm Mpc}$, which is similar to the scale of galaxy clusters today.  This means that it is possible for an environmentally dependent dilaton field to simultaneously obey local constraints and have an non-negligible effect on the formation of large scale cosmological structures.  We investigate this possibility further in the following section.

\section{Linear Structure Formation}\label{sec:Struct}
We found above that the cosmological range, $\lambda_{\rm cos}$, of the dilaton mediated fifth force was only constrained to be $\lesssim 0.5 - 2\,{\rm Mpc}$ today, depending on the value of $\lambda$.  Meanwhile the strength of this force (relative to gravity), $\alpha_{\rm cos}$, on scales smaller than $\lambda_{\rm cos}$ was found to lie in the range $0.04 - 0.33$,  for $\lambda_{} \gtrsim 10$, $\alpha_{\rm cos} \gtrsim 0.3$ and $\lambda_{\rm cos} \gtrsim 1\,{\rm Mpc}$. For such values of $\lambda$ we would, in the cosmological background, have a fifth force of strength similar to that of gravity propagating over a range that could be as large as that of the typical length scales of clusters of galaxies (i.e. $O(1)\,{\rm Mpc}$). Such a force would alter the formation of large scale structures in a manner that could be detected by ongoing and future galaxy surveys.
\begin{figure*}[tbh]
\begin{center}
\includegraphics[width=7.5cm]{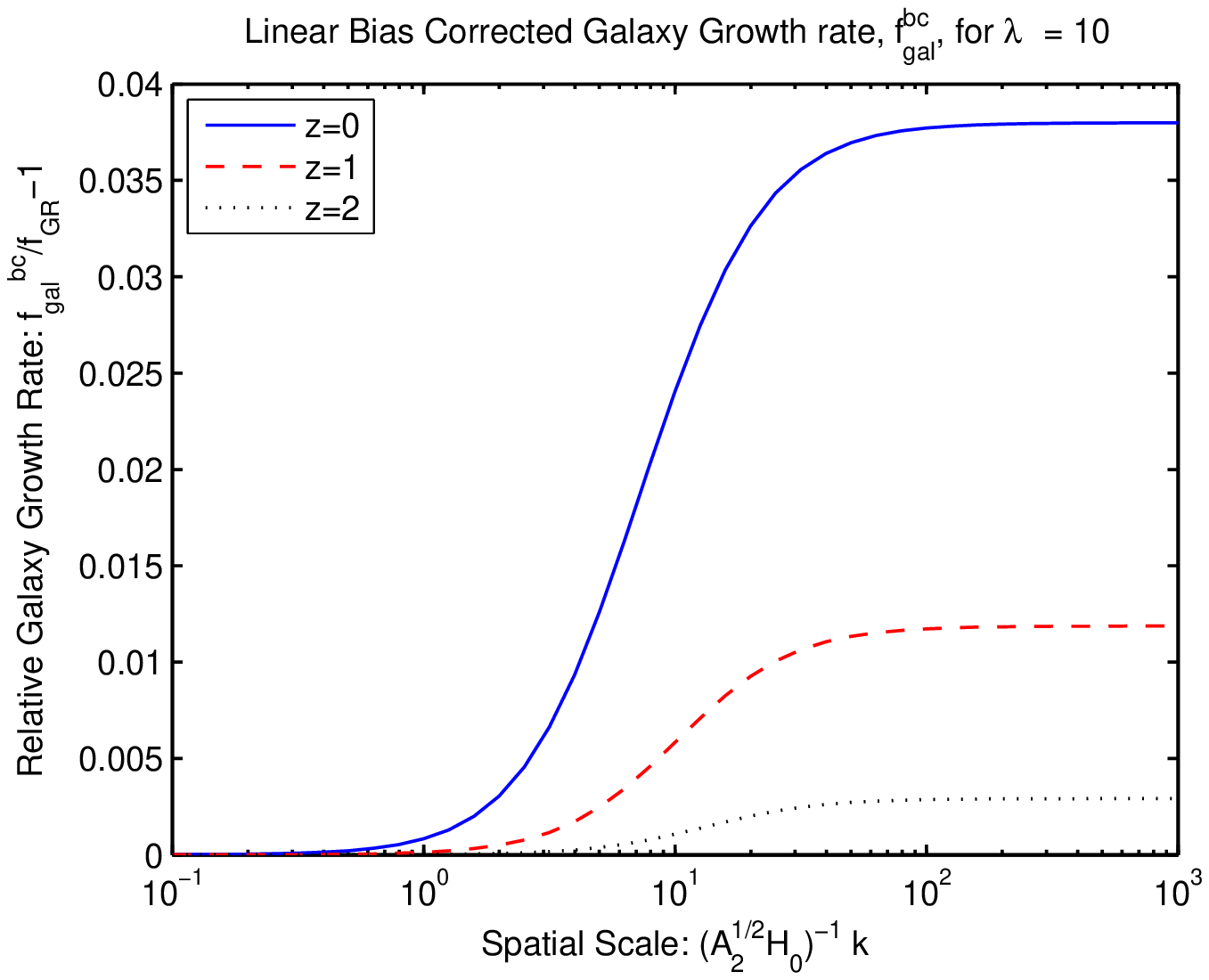}
\includegraphics[width=7.5cm]{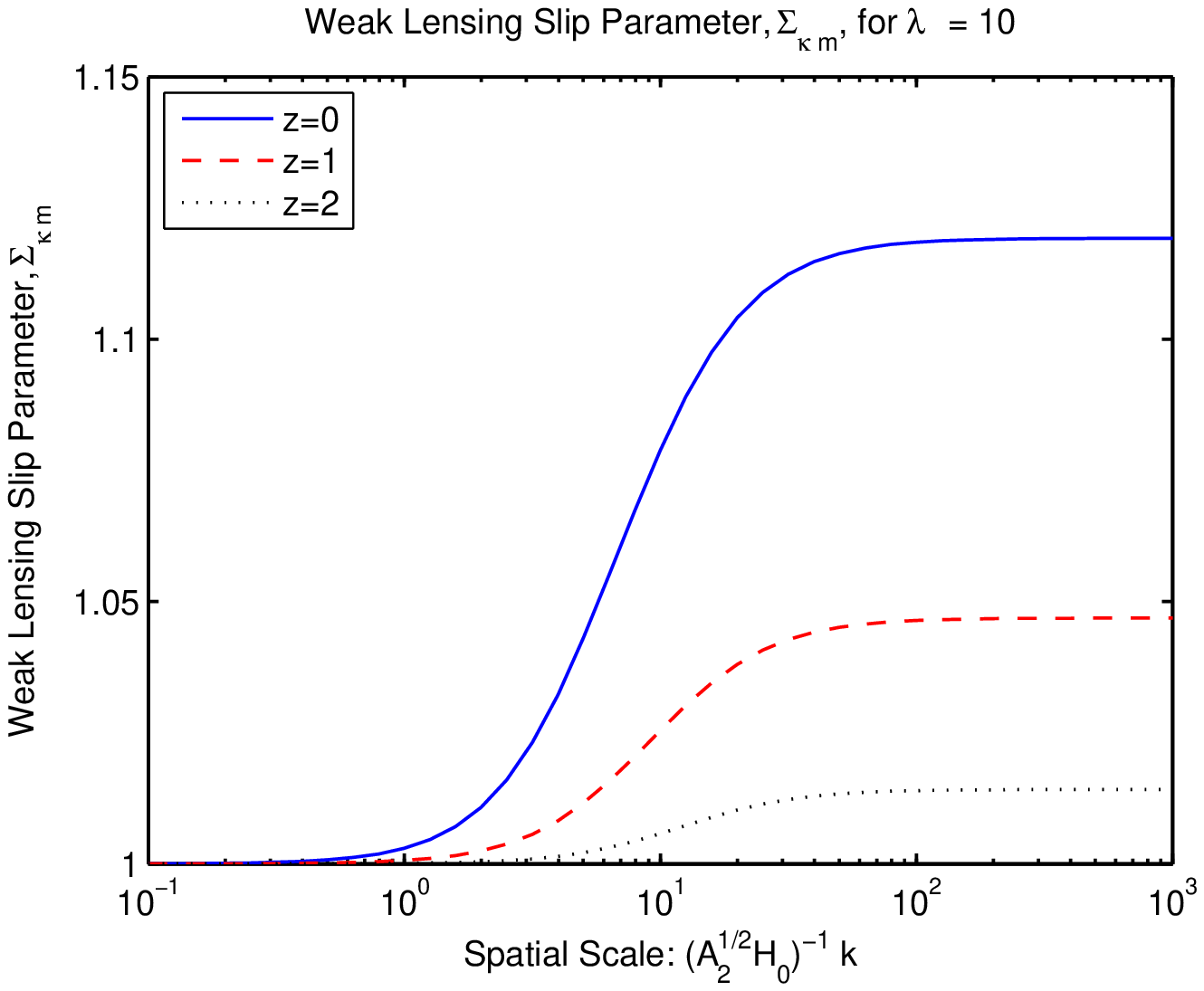}
\includegraphics[width=7.5cm]{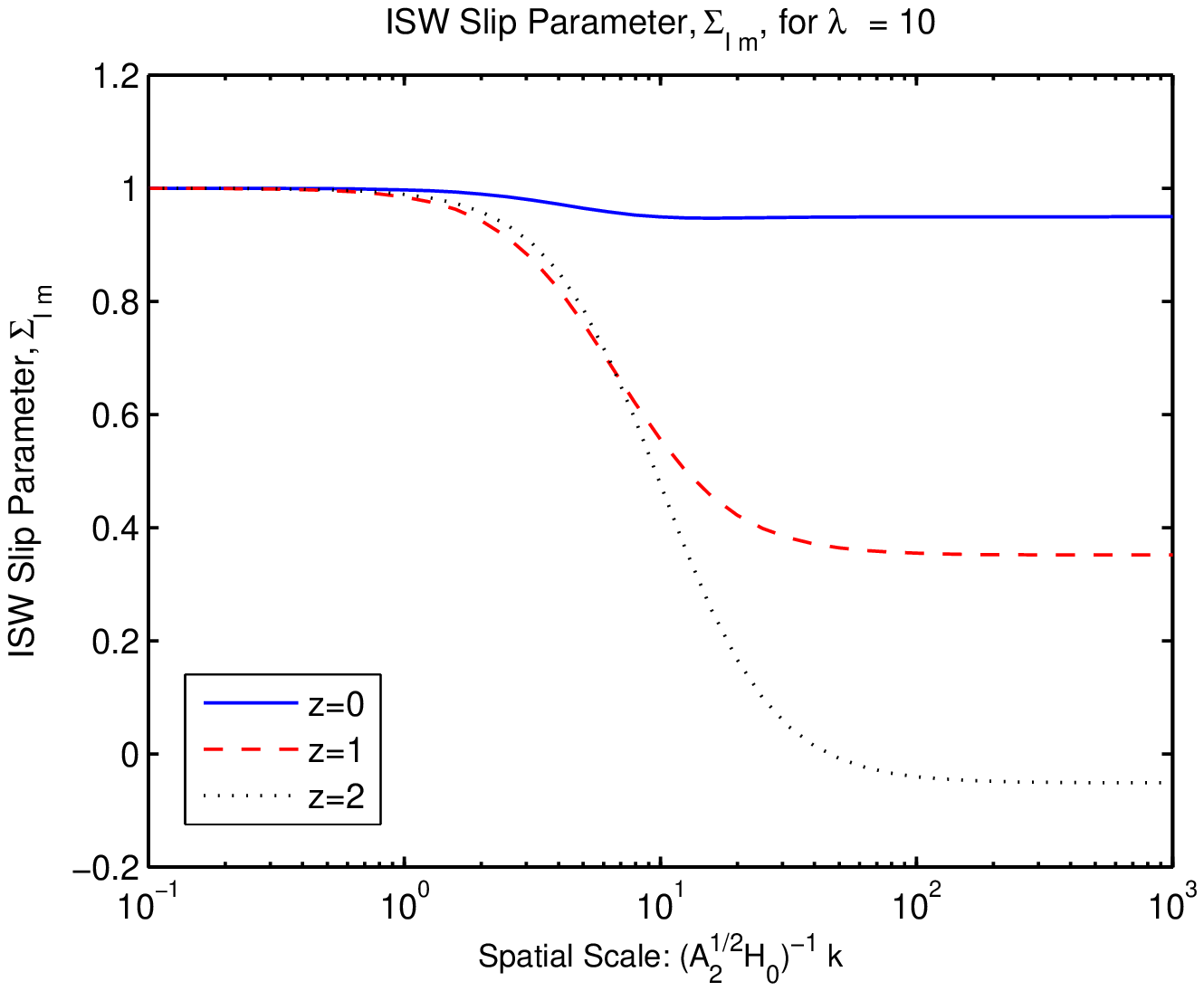}
\includegraphics[width=7.5cm]{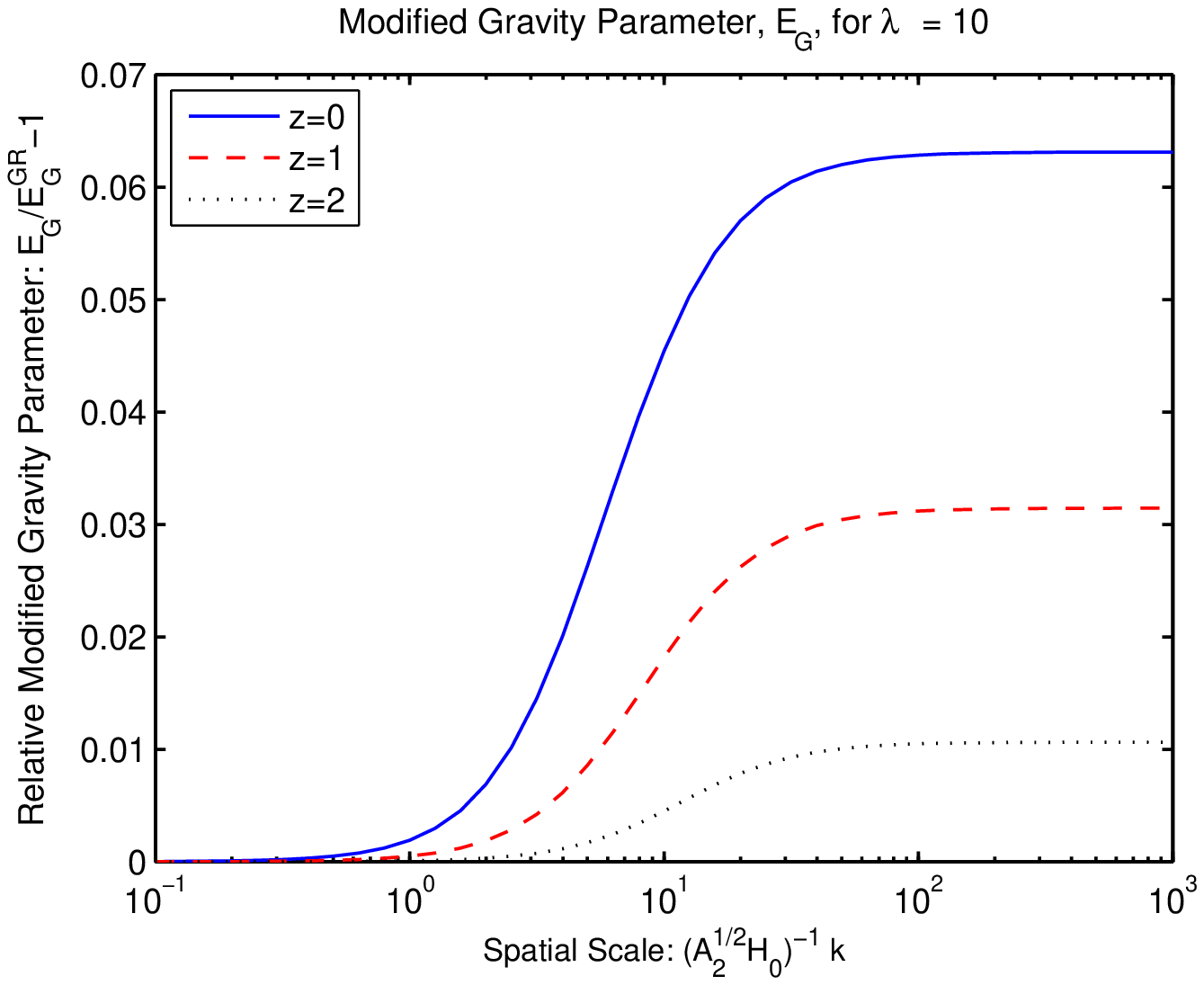}
\caption[]{Effect of the dilaton field on large scale structure formation.  From top to bottom the plots show: the predicted deviation of the (linear bias corrected) galaxy growth, $f_{\rm gal}^{\rm bc}$, from its GR value, $f_{\rm GR}$, the predicted value of the slip parameter, $\Sigma_{\rm kappa m}$, extrapolated from weak lensing measurements, the predicted slip parameter extrapolated from ISW measurements, $\Sigma_{\rm I m}$, and the relative deviation of the modified gravity parameter, $E_{\rm G}$, from its GR value.  These plots are for $\lambda = 10$, and show the values of $f_{\rm gal}^{\rm bc}$, $\Sigma_{\rm kappa m}$ and $\Sigma_{\rm Im}$ at the present day, $z=0$ (solid blue line), $z=1$ (dashed red line) and $z=2$ (dotted black line) for different values of the inverse spatial scale, $k$.  We see that the largest deviations from GR are found in $\Sigma_{\rm Im}$ particularly at $z\sim 1-2$ and the smallest in the growth rate.  This is because galaxies are effectively decoupled from the dilaton mediated fifth force whereas  $\Sigma_{\rm I m}$ as well as  $\Sigma_{\rm \kappa m}$ and $E_{G}$ directly probe the growth rate of the large scale dark matter perturbation which feels the unsuppressed fifth force.
\label{fig:struct1}}
\end{center}
\end{figure*}
In this section we detail the effect of our dilaton model on the structure formation in the linear regime.  The environmentally dependent nature of the model makes the study of non-linear structures considerably more complicated than in the standard case. We therefore postpone a detailed analysis of non-linear structure formation in this model to future work.

We define the Einstein frame energy momentum tensor $T^{\mu}_{\rm m}{}_{\nu} = A^3 \tilde{T}^{\mu}{}_{\nu}$ where  $\alpha = [(\ln A)_{,\phi}]^2 $.  The full field equations are then:
\be
\square \varphi &=& \frac{\kappa_4^2}{2k(\phi)}\left[ -V(\phi) - A^{\prime}(\phi) T^{\mu}_{\rm m}{}_{\mu}\right. \\ && \left. + 4\beta(\phi) V(\phi)\right]] \equiv V_{{\rm eff},\varphi}(\phi,T^{\mu}_{\rm m}{}_{\mu}), \nonumber \\
\nabla_{\mu}(A(\phi)T^{\mu}_{\rm m}{}_{\nu}) &=& T^{\mu}_{\rm m}{}_{\mu} \nabla_{\nu}A(\phi), \\
G^{\mu}{}_{\nu} &=& A(\phi)\kappa_4^2 T^{\mu}_{\rm m}{}_{\nu} - \kappa_4^2 V(\phi) \delta^{\mu}{}_{\nu} \\ &&+ 2\nabla_{\nu}\varphi \nabla^{\mu}\varphi - \delta^{\mu}{}_{\nu} (\nabla \varphi)^2.\nonumber
\ee
where $\dd \varphi = k(\phi)\dd \phi$ and we have defined an effective potential:
$$
V_{\rm eff}(\phi, T^{\mu}_{\rm m}{}_{\mu}) = V(\phi)-A(\phi)T^{\mu}_{\rm m}{}_{\mu}.
$$
The matter content is taken to be pressure-less dust $T^{\mu\nu}_{\rm m} = \rho u^{\mu}u^{\nu}$; $u^{\mu}u_{\mu} = -1$.   We consider the background FRW cosmology and denote background quantities by an over-bar, i.e. the background matter density is $\bar{\rho}$.  Hence:
\be
H^2 = \left(\frac{\dot{a}}{a}\right)^2 &=& \frac{\kap^2}{3}\left[A(\bar{\phi})\bar{\rho} + V(\bar{\phi})\right] + \frac{\dot{\bar{\varphi}}^2}{3}, \\
-\ddot{\bar{\varphi}}-3H\dot{\bar{\varphi}} &=& V_{\rm eff, \varphi}(\bar{\phi}; \bar{\rho}), \\
\bar{\rho} &\propto& a^{-3}.
\ee
The background mass for $\varphi$ is given by $m^2_\varphi = V_{{\rm eff}, \varphi \varphi)}(\bar{\phi};\bar{\rho})$.
Previously we found that the requirement $A_{2} \gg 1$, ensures that $m^2_{\varphi} \gg H^2$, and so $\bar{\varphi}$ (or equivalently $\bar{\phi}$) lies very close to the minimum of $V_{\rm eff}$.  We define $\bar{\phi}_{\rm min}$ by $V_{\rm eff,\phi}(\bar{\phi}_{\rm min},\bar{\rho}) = 0$, and then $\bar{\phi} \approx \bar{\phi}_{\rm min}$. With $A_2 \gg 1$, $m_{\varphi}^2 \gg H^2$, we also have $\vert\dot{\bar{\phi}}\vert \ll H$, and $A(\bar{\phi}) \approx 1$ which we set to $1$, and can take $\kap^2 V(\bar{\phi}) \approx \Lambda$. Here $\Lambda$ is the effective value of the cosmological constant.  Thus:
\be
H^2 &\approx& \frac{\kap^2}{3}\bar{\rho}  + \frac{\Lambda}{3}, \\
\bar{\beta} &\equiv& \beta(\bar{\phi}) = (\ln A(\bar{{\phi}}))^{\prime} \\ &\approx& \frac{\Lambda}{\kap^2 \bar{\rho}+4\kap^2 \Lambda} =  \frac{1-\Omega_m}{4-3\Omega_m}.\nonumber
\ee
The mass-squared, $m^2_{\varphi}$, is then given by Eq. \eqref{eq:cosmomass}.   We define $\bar{\alpha} = \beta^2(\bar{\phi})/k^2(\bar{\phi}) = \bar{\beta}^2/(3\bar{\beta}^2 + \lambda_{}^{-2})$.

We now focus on the growth of linear perturbations in the measured matter density $A(\phi)\rho$ around $A(\bar{\phi})\bar{\rho}$. Linearity here implies that  $\delta_m = \delta (A(\phi)\rho)/A(\bar{\phi})\bar{\rho}$, $\vert \delta_{\rm m} \vert \ll 1$. In addition to the usual gravitational force, linear perturbations now feel an additional fifth force with range $\lambda_{\rm cos} = 1/m_{\varphi}$ and strength $\bar{\alpha}$.  Since $m_{\varphi}^2 \gg H^2$, the evolution of the background cosmology is, however, $\Lambda$CDM to a very good approximation.

Galaxies represent non-linear perturbations in the matter density and have densities much greater than $\bar{\rho}$ and the matter coupling, $\alpha$ decreases as $1/\rho^2$ as $\rho$ increases.   Compatibility with local tests requires that the fifth force coupling inside our galaxy be greatly suppressed compared with its cosmological value. It is feasible that in smaller galaxies, the fifth force would be less suppressed, however for our purposes here we assume that on average $\alpha$ inside galaxies is much smaller than $\bar{\alpha}$ and so treat galaxies as being essentially uncoupled to the dilaton fifth force.

Galaxies are often used as observational tracers of the linear CDM perturbation.  The latter feels  both gravity and the fifth force whilst the former only evolves under gravity.   This is very similar to the scenario we considered in Ref. \cite{Brax:2009ab}. Using the calculations presented in Ref. \cite{Brax:2009ab}, and moving to Fourier space, we define $\delta_{\rm g}(k)$ and $\delta_{\rm m}(k)$ ($\delta_A$ and $\delta_B$ respectively in Ref. \cite{Brax:2009ab}) to be respectively the linear density perturbations with co-moving wave-number $k$ in the average density of galaxies and in the average density of all pressure-less matter.  Using Ref. \cite{Brax:2009ab} and with $p = \ln a$, we have:
\be
\delta_{{\rm g},pp}(k) &+& \left[2-\frac{3\Omega_m}{2}\right]\delta_{{\rm g},p}(k) = \frac{3}{2}\Omega_m \delta_{\rm m}, \\
\delta_{{\rm m},pp}(k) &+& \left[2-\frac{3\Omega_m}{2}\right]\delta_{{\rm m},p}(k) = \frac{3}{2}\left[1\right. \\ && \left. + \alpha_{\rm eff}(\Omega_m,a m_{\varphi}/k) \right]\Omega_m \delta_{\rm m}, \nonumber\\
\alpha_{\rm eff}(\Omega_{\rm m},x) &=& \frac{\bar{\alpha}(\Omega_{\rm m})}{1+x^2}, \\
\alpha(\Omega_m) &=& \left[3 + \lambda^{-2}\left(4+\frac{\Omega_{\rm m0}}{a^3(1-\Omega_{\rm m0})}\right)^{2}\right]^{-1},
\ee
where $\Omega_{\rm m0}$ and $H_0$ are the values of $\Omega_{\rm m}$ and $H$ today when $a=1$ ($p=0$).

The observables are the growth rate $f_{\rm gal} = \dd (\ln \delta_{\rm g}) /\dd \ln a$ , the slip functions, $\Sigma_{\kappa m}$ and $\Sigma_{\kappa I}$ measured by weak-lensing and  Integrated Sachs-Wolfe (ISW) effect measurements respectively, and finally the indicator of modified gravity $E_G$. The former two are defined in terms of the two metric potentials $\Phi$ and $\Psi$:
\begin{equation}
ds^2=a^2(\eta)\left ( -(1+2\Psi)d\eta^2+ (1-2\Phi)dx^2\right )
\end{equation}
Weak lensing measures $\Phi+\Psi$ and the ISW effect is proportional to $\dot{\Phi}+\dot{\Psi}$.  Note that in this model $\Phi=\Psi$ in the Einstein frame.  $\Sigma_{\kappa m}$ and $\Sigma_{\kappa I}$ are then given by:
\be
k^2(\Phi+\Psi) &=& -8\pi G a^2 \bar{\rho} \Sigma_{\kappa m} D_{\rm GR}\delta_{\rm i}, \\
H^{-1} k^2 (\dot{\Phi}+\dot{\Psi})&=& -8\pi G a^2 \bar{\rho} \Sigma_{\kappa m} (f_{\rm GR}-1) D_{\rm GR}\delta_{\rm i},
\ee
where $\delta_{i}$ is the primordial density perturbation (measured from the CMB), and $D_{\rm GR}$ is the growth factor in GR; $\delta_{\rm m} = D_{\rm GR}\delta_{i}$, $f_{\rm GR} = \dd \ln D_{\rm GR}/\dd \ln a$ is the GR growth rate.  In this model $\delta_{\rm m} = D_{\rm m} \delta_{i}$ and so $\Sigma_{\kappa m} = D_{\rm m}/D_{\rm GR}$ and $\Sigma_{\kappa I} = (f_{\rm m}-1)\Sigma_{\kappa m}/(f_{\rm GR}-1)$; $f_{\rm m} = \dd \ln D_{\rm m} / \dd \ln a = \dd \ln \delta_{\rm m} / \dd \ln a$.
\begin{figure*}[tbh]
\begin{center}
\includegraphics[width=7.5cm]{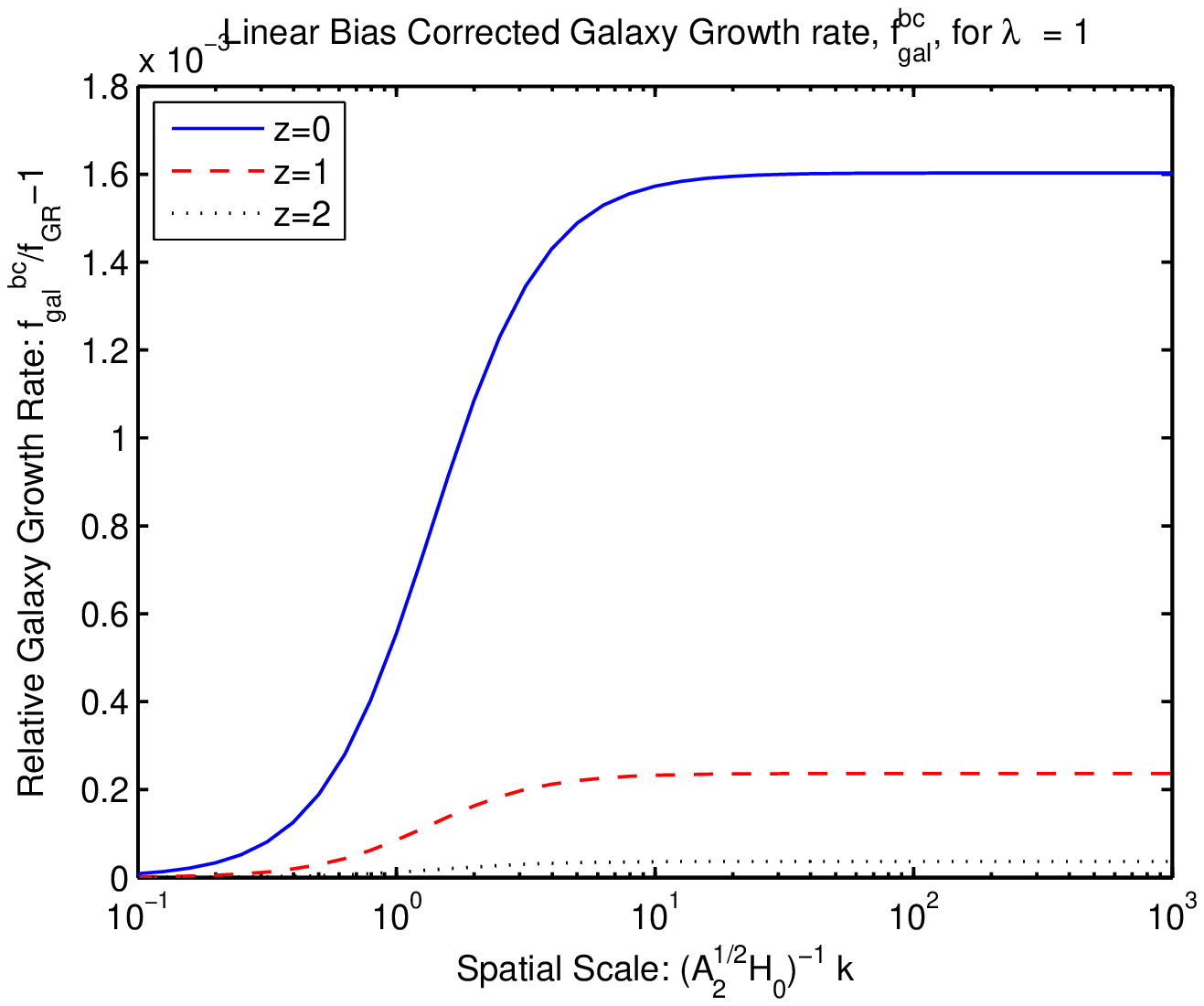}
\includegraphics[width=7.5cm]{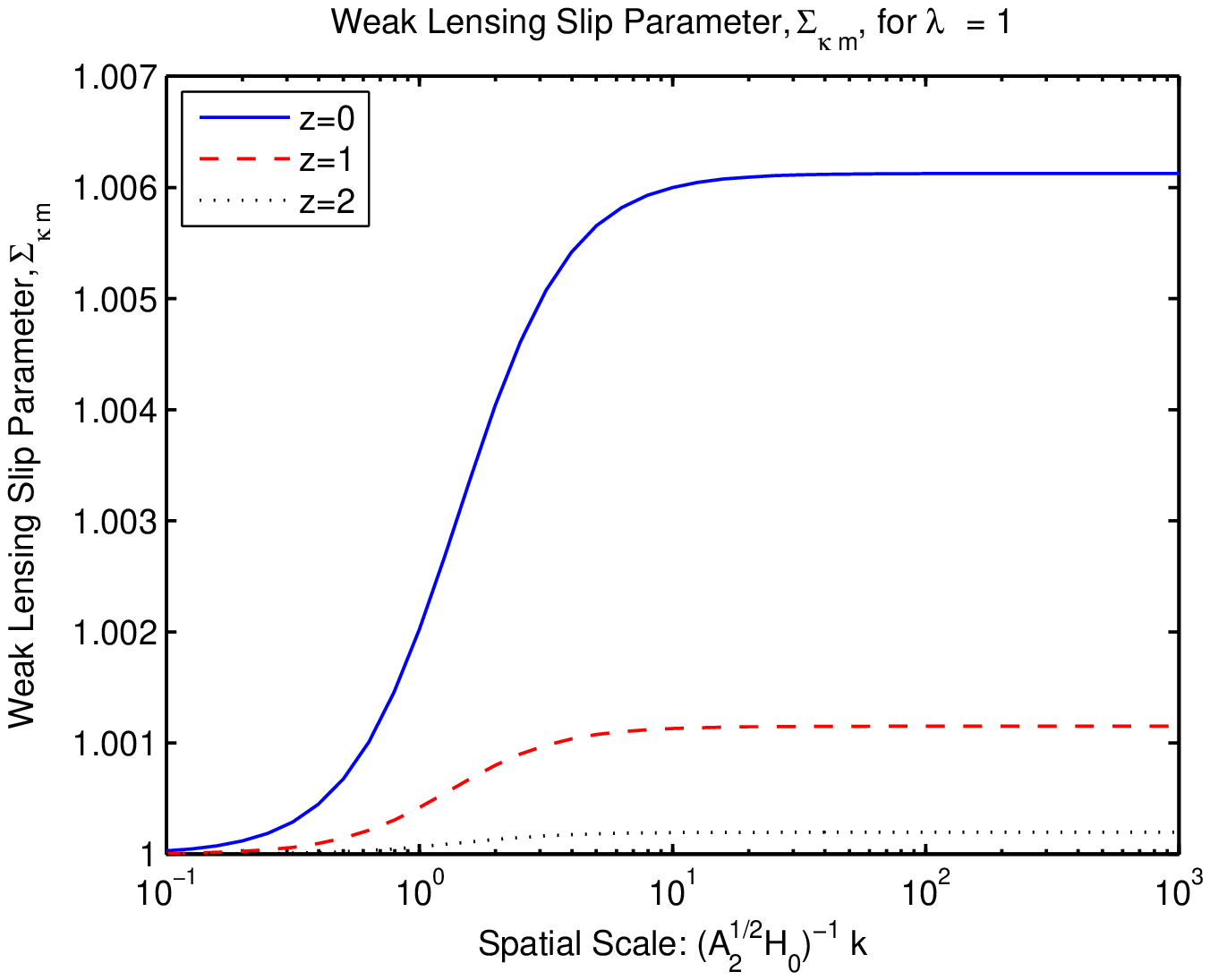}
\includegraphics[width=7.5cm]{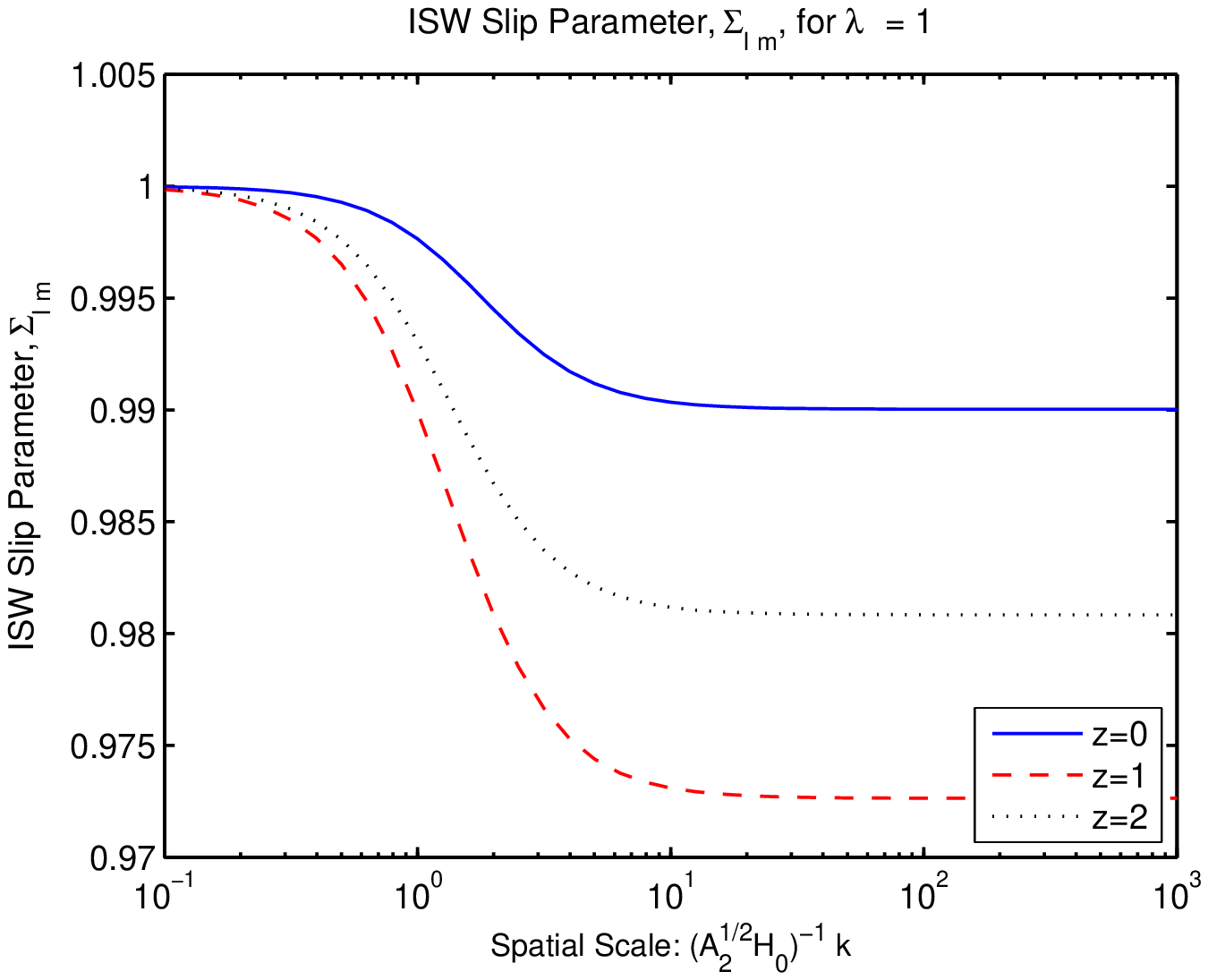}
\includegraphics[width=7.5cm]{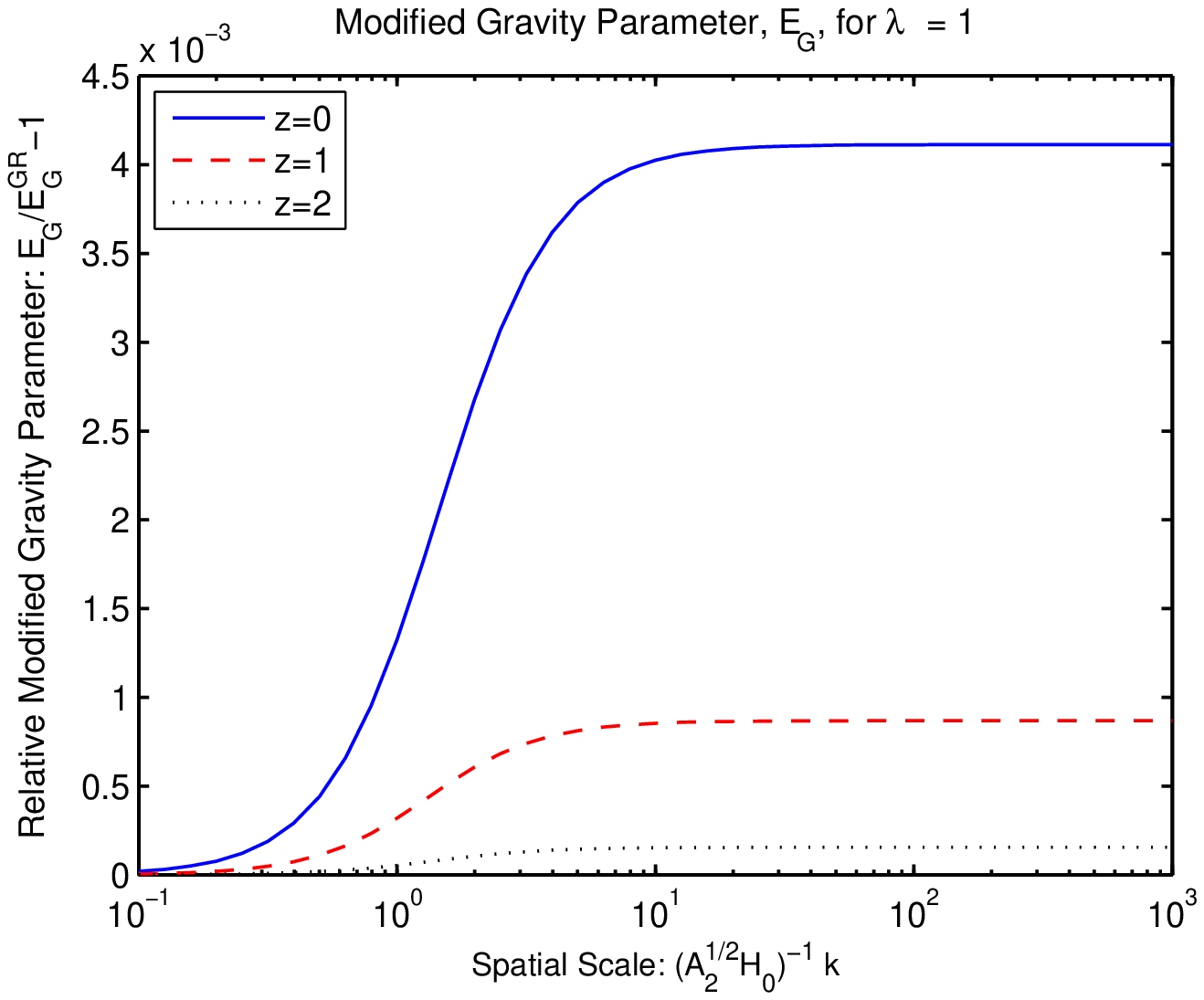}
\caption[]{Effect of the dilaton field on large scale structure formation.  From top to bottom the plots show: the predicted deviation of the (linear bias corrected) galaxy growth, $f_{\rm gal}^{\rm bc}$, from its GR value, $f_{\rm GR}$, the predicted value of the slip parameter, $\Sigma_{\rm \kappa m}$, extrapolated from weak lensing measurements, the predicted slip parameter extrapolated from ISW measurements, $\Sigma_{\rm I m}$, and the relative deviation of the modified gravity parameter, $E_{\rm G}$, from its GR value. These plots are for $\lambda = 1$, and show the values of $f_{\rm gal}^{\rm bc}$, $\Sigma_{\rm \kappa m}$ and $\Sigma_{\rm Im}$ at the present day, $z=0$ (solid blue line), $z=1$ (dashed red line) and $z=2$ (dotted black line) for different values of the inverse spatial scale, $k$.   We see that the largest deviations from GR are found in $\Sigma_{\rm Im}$ particularly at $z\sim 1-2$ and the smallest in the growth rate.  This is because galaxies are effectively decoupled from the dilaton mediated fifth force whereas  $\Sigma_{\rm I m}$ as well as  $\Sigma_{\rm \kappa m}$ and $E_{G}$ directly probe the growth rate of the large scale dark matter perturbation which feels the unsuppressed fifth force.   We note that the deviations from GR are much smaller for $\lambda = 1$ than there are for $\lambda =10$ (see FIG. \ref{fig:struct1}).  This is because for $\lambda \gg 1$, the cosmological fifth force coupling, $\alpha_{\rm cos}$, $\approx 1/3$ today whereas for $\lambda=1$ it is only $\approx 0.04$.
\label{fig:struct2}}
\end{center}
\end{figure*}

We also define the linear bias corrected galaxy growth rate.  This is defined by assuming that $\delta_{\rm g} = D_{\rm gal}^{bc} \delta_{i} + \Delta_{0}$ where $\delta_{i}$ is the initial Gaussian perturbation, $D_{\rm gal}^{bc}$ is the bias corrected growth factor for galaxies, and $\Delta_{0}$ is the source of the bias.  We then have that $\delta_{\rm g}^{bc} = D_{\rm gal}^{bc}\delta_{i} = b_{\rm lin}^{-1}(\delta_{g})\delta_{g}$ where $b_{\rm lin}^{-1} = 1-\Delta_{0}/\delta_{g}$.  $\Delta_{0}$ and hence the linear bias  is estimated directly from galaxy surveys using higher order statistics.    We define $f_{\rm gal}^{\rm bc} = \dd \ln \delta_{\rm g}^{\rm bc}/\dd \ln a$. In the absence of any deviations from GR, $f_{\rm gal}^{bc} = f_{\rm GR} \approx \Omega_{\rm m}^{0.545}$, and $\Sigma_{\kappa m} =\Sigma_{\kappa I} = 1$.

 Finally, we consider the modified gravity sensitivity parameter defined in Ref. \cite{dodelson}:
\be
E_{\rm G} = \frac{k^2 (\Psi+\Phi)}{-3H_0^2 a^{-1}\theta} , \nonumber
\ee
where $\theta = -\dot{\delta}_{\rm gal}/H = -\dd \delta_{\rm gal}/\dd \ln a$, and so $\theta = -f_{\rm gal}^{\rm bc}D_{\rm gal}^{bc}\delta_{i}$ and with $a=1$ today:
\be
E_{\rm G} = \frac{\Omega_{\rm m0} D_{\rm m}}{f_{\rm gal}^{\rm bc}D_{\rm gal}^{bc}}
\ee
In GR, $E_{\rm G}^{(\rm GR)} = \Omega_{\rm m0}/f_{\rm GR} \approx \Omega_{\rm m0} \Omega_{\rm m}^{-0.545}$.
In all cases, we find that gravity is modified below a certain red-shift $z_\ast$ approximately given by $(1+z_\ast) \sim \lambda^{1/3}$. Hence gravity is only modified at low redshift for scales which are small enough, inside the Compton wavelength  of the dilaton on cosmological scales.

 FIGs \ref{fig:struct1} and \ref{fig:struct2}  show $f_{\rm gal}^{\rm bc}/f_{\rm GR}-1$, $\Sigma_{\rm \kappa m}$ and $\Sigma_{\rm I m}$ and $E_{\rm G}/E_{\rm G}^{\rm (GR)}-1$ for $\lambda =10$ and $\lambda =1$ respectively and for different values of redshift, $z$, and inverse spatial scale $k$.  We see that the largest deviations from GR occur on scales $k > A_{2}^{1/2}H_0$ (roughly $k > m_{\rm cos}$), and at late times.  The deviations from GR also scale with $\alpha_{\rm cos}$ and hence with $\lambda$; for $\lambda \gg 1$, $\alpha_{\rm cos}(z=0) \approx 1/3$ whereas for $\lambda =1$, $\alpha_{\rm cos}(z=0) \approx 0.04$.  Additionally, $\Sigma_{I m}$, $\Sigma_{\kappa m}$ and $E_{\rm G}$ display more pronounced deviations from their GR values than does $f_{\rm gal}^{\rm bc}$.   This is because galaxies are effectively decoupled from the dilaton mediated fifth force whereas both $\Sigma_{I m}$, $E_{\rm G}$ and $\Sigma_{\rm Im}$ directly probe the growth rate of the large scale dark matter perturbation which feels the unsuppressed fifth force. For $\lambda = 10$, the largest deviations from GR occur in $\Sigma_{Im}$ at redshifts $z\sim 1-2$, where $1-\Sigma_{Im} \sim O(1)$.  In other parameters the greatest deviations occur today ($z=0$), with $f_{\rm gal}^{\rm bc}$, $\Sigma_{\kappa m}$ and $E_{\rm G}$ deviating from their GR values by $\approx 3.7\%$, $12\%$ and $6\%$ respectively.   We note that these deviations are within present constraints but should be detectable by future surveys of large scale structure.

\section{Conclusions}

We have presented new results on dilaton models in the strong coupling regime. Our analysis combines a runaway potential as suggested in \cite{ven1} with a coupling to matter with a vanishing minimum, as investigated in \cite{Damour:1994zq}.  In the absence of any potential term, the coupling to matter is enough to drive the dilaton to its minimum cosmologically \cite{Damour:1994zq}. As a result, the dilaton would evade gravitational tests on the non-existence of fifth forces. This result is jeopardised by the runaway dilatonic potential, in particular if the dilaton plays the role of dark energy. In this paper we have focussed on the dilaton as a candidate for dark energy and shown how this can be compatible with gravitational tests of gravity, but could lead to observational tests at large scales.

We have shown that for models where the string scale is lower than the Planck scale, the Damour-Polyakov mechanism, whereby the coupling to matter vanishes dynamically, is at play only locally in the presence of large enough over-densities such as the one present in galaxies. On larger cosmological scales, the fifth force is not suppressed implying the presence of a significant modification of gravity. We found that this effect could be relevant on galaxy cluster scales where the growth of structures would be affected. As a result, the future galaxy survey could give stringent constraints on dilaton models and have the potential to indicate the possible existence of a dilaton in the strong coupling regime. We also expect that the non-linear growth of structures would also be affected by the presence of a dilaton. This is left for future work.

\acknowledgments We are grateful to G. Veneziano and M. Gasperini for
  discussions. We also thank G. Veneziano for comments on the
  draft. DS is supported by STFC. CvdB and ACD are partly supported by
  STFC.


\begin{thebibliography}{99}

\bibitem{various}
E.~J.~Copeland, M.~Sami and S.~Tsujikawa,
 Int.\ J.\ Mod.\ Phys.\  D {\bf 15} (2006) 1753 [arXiv:hep-th/0603057],
  S.~M.~Carroll,
  Living Rev.\ Rel.\  {\bf 4} (2001) 1
  [arXiv:astro-ph/0004075],
  P.~J.~E.~Peebles and B.~Ratra,
  Rev.\ Mod.\ Phys.\  {\bf 75}, 559 (2003)
  [arXiv:astro-ph/0207347],
  Ph.~Brax,  arXiv:0912.3610.

\bibitem{DGP}
G.~R.~Dvali, G.~Gabadadze and M.~Porrati,
  Phys.\ Lett.\  B {\bf 485} (2000) 208
  [arXiv:hep-th/0005016].

\bibitem{Vainshtein}
  A.~I.~Vainshtein,
  Phys.\ Lett.\  B {\bf 39} (1972) 393.

\bibitem{chameleon1}
  J.~Khoury and A.~Weltman,
  Phys.\ Rev.\  D {\bf 69} (2004) 044026
  [arXiv:astro-ph/0309411],
  J.~Khoury and A.~Weltman,
  Phys.\ Rev.\ Lett.\  {\bf 93} (2004) 171104
  [arXiv:astro-ph/0309300].

\bibitem{chameleon2}
P.~Brax, C.~van de Bruck, A.~C.~Davis, J.~Khoury and A.~Weltman,
  Phys.\ Rev.\  D {\bf 70} (2004) 123518
  [arXiv:astro-ph/0408415].

\bibitem{chameleon3}
D.~F.~Mota and D.~J.~Shaw,
  Phys.\ Rev.\ Lett.\  {\bf 97} (2006) 151102
  [arXiv:hep-ph/0606204],
  D.~F.~Mota and D.~J.~Shaw,
  Phys.\ Rev.\  D {\bf 75} (2007) 063501
  [arXiv:hep-ph/0608078].


\bibitem{f(R)chameleon}
A.~W.~Brookfield, C.~van de Bruck and L.~M.~H.~Hall,
  Phys.\ Rev.\  D {\bf 74} (2006) 064028
  [arXiv:hep-th/0608015],
  T.~Faulkner, M.~Tegmark, E.~F.~Bunn and Y.~Mao,
  Phys.\ Rev.\  D {\bf 76} (2007) 063505
  [arXiv:astro-ph/0612569],
  W.~Hu and I.~Sawicki,
  Phys.\ Rev.\  D {\bf 76} (2007) 064004
  [arXiv:0705.1158 [astro-ph]],
  P.~Brax, C.~van de Bruck, A.~C.~Davis and D.~J.~Shaw,
  Phys.\ Rev.\  D {\bf 78} (2008) 104021
  [arXiv:0806.3415 [astro-ph]].


\bibitem{Damour:1994zq}
  T.~Damour and A.~M.~Polyakov,
  Nucl.\ Phys.\  B {\bf 423}, 532 (1994)
  [arXiv:hep-th/9401069].

\bibitem{ven1} M. Gasperini, F. Piazza and  G. Veneziano
Phys.Rev. D65 (2002) 023508, [arXiv:gr-qc/0108016].

\bibitem{ven2}
T. Damour, F. Piazza and  G. Veneziano, Phys.Rev. D66 (2002) 046007, [arXiv:hep-th/0205111 ]

\bibitem{ven3}
 Thibault Damour, Federico Piazza, Gabriele Veneziano
Phys.Rev.Lett. 89 (2002) 081601, [arXiv:gr-qc/0204094].


\bibitem{Green:1987sp}
  M.~B.~Green, J.~H.~Schwarz and E.~Witten, {\it Superstring Theory}, (Cambridge University Press, Cambridge, England, 1987).

\bibitem{brustein}
R.~Brustein, G.~Dvali and G.~Veneziano,
  JHEP {\bf 0910}, 085 (2009)
  [arXiv:0907.5516 [hep-th]].

\bibitem{Bertotti:2003rm}
  B.~Bertotti, L.~Iess and P.~Tortora,
  Nature {\bf 425}, 374 (2003).

 \bibitem{Will:2001mx}
  C.~M.~Will,
  Living Rev.\ Rel.\  {\bf 4} (2001) 4
  [arXiv:gr-qc/0103036].

\bibitem{Schlamminger:2007ht}
  S.~Schlamminger, K.~Y.~Choi, T.~A.~Wagner, J.~H.~Gundlach and E.~G.~Adelberger,
  Phys.\ Rev.\ Lett.\  {\bf 100}, 041101 (2008)
  [arXiv:0712.0607 [gr-qc]].

\bibitem{Dent:2008gu}
  T.~Dent,
  Phys.\ Rev.\ Lett.\  {\bf 101}, 041102 (2008)
  [arXiv:0805.0318 [hep-ph]].

\bibitem{Xue:2008se}
  X.~X.~Xue {\it et al.}  [SDSS Collaboration],
  Astrophys.\ J.\  {\bf 684}, 1143 (2008)
  [arXiv:0801.1232 [astro-ph]].

\bibitem{Amsler:2008zzb}
  C.~Amsler {\it et al.}  [Particle Data Group],
  Phys.\ Lett.\  B {\bf 667}, 1 (2008).

\bibitem{Brax:2009ab}
  P.~Brax, C.~van de Bruck, A.~C.~Davis and D.~Shaw,
  JCAP {\bf 1004} (2010) 032
  [arXiv:0912.0462 [astro-ph.CO]].


\bibitem{dodelson}
 P.~Zhang, M.~Liguori, R.~Bean and S.~Dodelson,
 Phys.\ Rev.\ Lett.\  {\bf 99} (2007) 141302
 [arXiv:0704.1932 [astro-ph]].

\end{thebibliography}
\end{document}